\documentclass[aps,prl,twocolumn,notitlepage]{revtex4-2}

\usepackage{amsmath}
\usepackage{amssymb}
\usepackage{amsfonts}
\usepackage{graphicx}
\usepackage{bm}
\usepackage{epstopdf}

\usepackage{hyperref} 

\usepackage{tensor}
\usepackage{tikz}
\usepackage{dsfont}

\definecolor{kdblue}{RGB}{0,65,145}
\definecolor{kblue}{RGB}{20,135,200}
\definecolor{kmblue}{RGB}{0,65,135}
\definecolor{klblue}{RGB}{95,190,235}
\definecolor{kdgray}{RGB}{124,124,124}

\definecolor{bpink}{RGB}{255,166,252}
\definecolor{bppink}{RGB}{255,118,216}
\definecolor{bpppink}{RGB}{255,89,199}

\definecolor{chv1red}{HTML}{BF1717}
\definecolor{chv1yellow}{HTML}{D9CF9C}
\definecolor{chv1blue}{HTML}{350FBF}
\definecolor{chv1lblue}{HTML}{4183D9}

\definecolor{chv2_1}{RGB}{229,229,219}
\definecolor{chv2_2}{RGB}{234,220,219}
\definecolor{chv2_3}{RGB}{231,205,208}
\definecolor{chv2_4}{RGB}{211,178,197}
\definecolor{chv2_5}{RGB}{187,120,155}

\definecolor{chv3pink}{HTML}{F252E8}
\definecolor{chv3ppink}{HTML}{A60F71}
\definecolor{chv3blue}{HTML}{275AF2}
\definecolor{chv3lblue}{HTML}{22CCF2}
\definecolor{chv3brown}{HTML}{BF5F0B}

\definecolor{tronblue}{HTML}{5BC2D9}
\definecolor{tronlblue}{HTML}{99F2F2}
\definecolor{trongreen}{HTML}{0FF207}

\definecolor{spider1}{HTML}{BDC5F2}

\hypersetup{
  colorlinks = true,    
  linkcolor = kblue,    
  citecolor  = bpppink,   
  urlcolor   = chv3lblue     
}

\begin{document}

\title{Effective Field Theory of a Noncollinear Altermagnet}

\author{Seungho Lee}
\affiliation{Department of Physics, Korea Advanced Institute of Science and Technology, Daejeon 34141, Republic of Korea}

\author{Se Kwon Kim}
\email{sekwonkim@kaist.ac.kr}
\affiliation{Department of Physics, Korea Advanced Institute of Science and Technology, Daejeon 34141, Republic of Korea}


\begin{abstract}
  We derive an effective field theory for a noncollinear altermagnet and magnons on top of the noncollinear ground state from an altermagnetic Heisenberg model. We obtain the ground-state phase diagram, revealing a noncollinear phase and four distinct collinear phases. The ground state of the noncollinear phase fully breaks the spin rotational symmetry, while the ground state of the collinear phases possesses unbroken $\mathrm{SO}(2)$ symmetry. The resulting effective field theory for the noncollinear phase is an $\mathrm{SO}(3)$ sigma model in which the magnonic excitation has three independent degrees of freedom and exhibits the $d$-wave-like anisotropic linear dispersion. We also discuss possible topological solitons, including $\mathbb{Z}_2$ vortices.
\end{abstract}

\maketitle

\emph{Introduction.}|Frustrated magnets, governed by competing interactions, have been extensively investigated due to their nontrivial ground states and topological excitations such as chiral soliton lattices, vortices, Skyrmions, and Hopfions~\cite{Dombre1989, Ulloa2016, Ochoa2018, Zarzuela2019a, Zarzuela2021, Zarzuela2025a, Batista2018, Schubring2020, Schubring2021, Naya2022, Lin2016, Sutcliffe2017, Harland2019, Speight2020, Lohani2019,  Pradenas2024, Tchernyshyov2024a, Pradenas2024a, Wernert2025, Pradenas2025}. Among the myriad of frustrated magnetic systems, noncollinear antiferromagnets have attracted increasing attention in recent years. In particular, noncollinear ground states of antiferromagnets in triangular, kagome lattices, and pyrochlore lattices, and their field theories have been studied in Refs.~\cite{Dombre1989, Ulloa2016, Batista2018, Pradenas2024, Tchernyshyov2024a, Pradenas2024a, Wernert2025, Pradenas2025}. The noncollinear spin configuration fully breaks the spin rotational symmetry and thereby its low-energy physics is parameterized by a $\mathrm{SO}(3)$ order parameter, while a collinear magnet is described by one unit vector field. Consequently, the number of magnons---the Nambu-Goldstone bosons associated with the broken spin-rotational symmetry---and the types of topological solitons, which include domain walls, vortices, and Skyrmions, differ from those of a conventional collinear magnet.
The interaction between the magnons and topological textures of the $\mathrm{SO}(3)$ order parameter has been studied in Ref.~\cite{Zarzuela2025a}.

Recently emerging unconventional antiferromagnets called altermagnets have gained significant interest due to their unique properties and potential applications in spintronics and magnonics~\cite{Smejkal2022, Smejkal2022a, Jiang2024, Chen2024g, Xiao2024, Bhowal2024}. Collinear altermagnets are characterized by a momentum-dependent spin splitting in their electronic and magnonic bands despite vanishing net magnetization. A Landau theory for collinear altermagnets with multipolar order parameter has been formulated in Ref.~\cite{McClarty2024a}. Concurrently, altermagnetism has been observed and theoretically studied in microscopic models~\cite{Gomonay2024a, Yershov2024b, Yershov2025, Jin2024b, Ferrari2024a, Consoli2025}. Refs.~\cite{Gomonay2024a, Yershov2024b, Yershov2025, Jin2024b} uses the continuum Lagrangian approach to describe the low-energy physics of collinear altermagnets. In the continuum theory, altermagnetic nature appears as a spatial anisotropy in the effective action. Although Refs.~\cite{Gomonay2024a, Yershov2024b, Yershov2025, Jin2024b} have focused only on the collinear phase, the microscopic models generally allow noncollinear phases due to the frustrating interactions. Hence, novel phenomena arising from the geometric frustration can be extended to the noncollinear phases of the altermagnetic models by analogy with noncollinear antiferromagnets.

In this Letter, we find a noncollinear phase with exact phase boundaries in an altermagnetic Heisenberg model in a bilayer square lattice~\cite{Gomonay2024a, Yershov2024b, Yershov2025, Jin2024b} and derive an $\mathrm{SO}(3)$ sigma model as its effective field theory. While noncollinear ordering of altermagnets has already been reported~\cite{Xiao2024, Chen2024g, Jiang2024}, an effective field theory has not been explicitly derived yet. Our work provides the first example of the effective field theory for noncollinear altermagnets. Moreover, previous $\mathrm{SO}(3)$ sigma models derived from microscopic Hamiltonians in magnetic systems have been restricted to triangular, kagome, and pyrochlore lattices. In this work, we extend the framework by deriving the $\mathrm{SO}(3)$ sigma model on a bilayer square lattice.

\begin{figure}[t]
  \includegraphics[width=1\columnwidth]{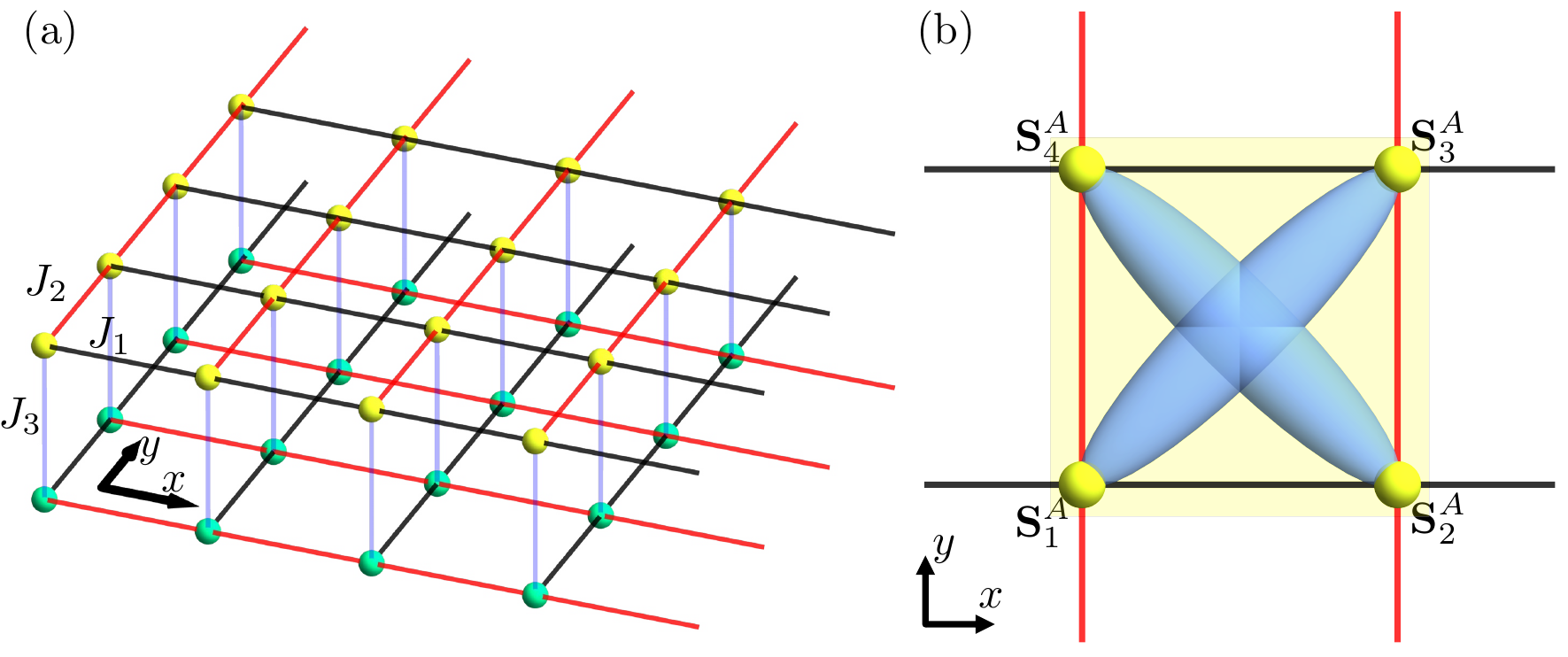}
  \caption{(a) The bilayer square lattice with alternating exchange interactions. The black, red, and blue lines represent the exchange interactions with the coefficient $J_1$, $J_2$, and $J_3$, respectively. The yellow (green) dots indicate spins at layer $A$ ($B$). (b) Pairs of spins $\{\mathbf{S}^A_1, \mathbf{S}^A_3\}$ and $\{\mathbf{S}^A_2, \mathbf{S}^A_4\}$ for producing the N\'eel order parameter in the unit cell of the layer $A$.}
  \label{fig:1}
\end{figure}

\emph{Model.}|We consider a minimal model of altermagnets, which contains the alternating Heisenberg exchange interaction on the bilayer square lattice~\cite{Jin2024b}. The Hamiltonian is given by
\begin{equation}\label{eq:ham}
\begin{aligned}
  H & = H_A + H_B + H_\mathrm{int}\,,
  \\
  H_A&= - \sum_{\mathbf{x}} \left[J_1\,\mathbf{S}^{A}_{\mathbf{x}}\cdot \mathbf{S}^{A}_{\mathbf{x}+a\hat{x}} +  J_2\,\mathbf{S}^{A}_{\mathbf{x}}\cdot \mathbf{S}^{A}_{\mathbf{x}+a\hat{y}}\right]\,,
  \\
  H_B&= - \sum_{\mathbf{x}} \left[J_2\,\mathbf{S}^{B}_{\mathbf{x}}\cdot \mathbf{S}^{B}_{\mathbf{x}+a\hat{x}} +  J_1\,\mathbf{S}^{B}_{\mathbf{x}}\cdot \mathbf{S}^{B}_{\mathbf{x}+a\hat{y}}\right]\,,
  \\
  H_{\mathrm{int}}&= - \sum_{\mathbf{x}} J_3\,\mathbf{S}^{A}_{\mathbf{x}}\cdot \mathbf{S}^{B}_{\mathbf{x}}\,,
\end{aligned}
\end{equation}
where the lattice site is defined by $\mathbf{x} = a(n\hat{x} + m\hat{y})$ with $n,m\in\mathbb{Z}$, $a$ is the lattice constant, and $\mathbf{S}^{A(B)}_{\mathbf{x}}$ is the unit-length spin at the site $\mathbf{x}$ in the layer $A(B)$ (see Fig.~\ref{fig:1}). Hereafter, we assume $J_1>0$ and $J_2 \leq J_1$. When $J_2 < 0$, this system generally is frustrated, and the ground state can therefore exhibit a noncollinear order. To analyze the noncollinear order, it is convenient to introduce the N\'eel fields and magnetization fields defined by
\begin{equation}
  \begin{aligned}
  \mathbf{S}_{1,3}^A = \mp \mathbf{N}_1^A + \mathbf{M}_1^A\,, \quad \mathbf{S}_{2,4}^A = \mp \mathbf{N}_2^A + \mathbf{M}_2^A\,,
  \\
  \mathbf{S}_{1,3}^B = \mp \mathbf{N}_1^B + \mathbf{M}_1^B\,, \quad \mathbf{S}_{2,4}^B = \pm \mathbf{N}_2^B + \mathbf{M}_2^B\,,
  \end{aligned}
\end{equation}
with the constraints $\mathbf{N}^I_i \cdot \mathbf{M}^I_i = 0$, where $I=A,B$ and $i=1,2$.
Here, $\mathbf{S}^A_{1,2,3,4}$ are defined as in Fig.~\ref{fig:1} (b) and $\mathbf{S}^B_{1,2,3,4}$ are defined similarly. In the continuum limit with these smooth variables, we can expand the Hamiltonian up to the quadratic order of derivatives $H\approx H^{(0)} + H^{(2)}$, with a vanishing linear term $H^{(1)} =0$~\cite{smalt}.

\begin{figure}[t]
  \includegraphics[width=1\columnwidth]{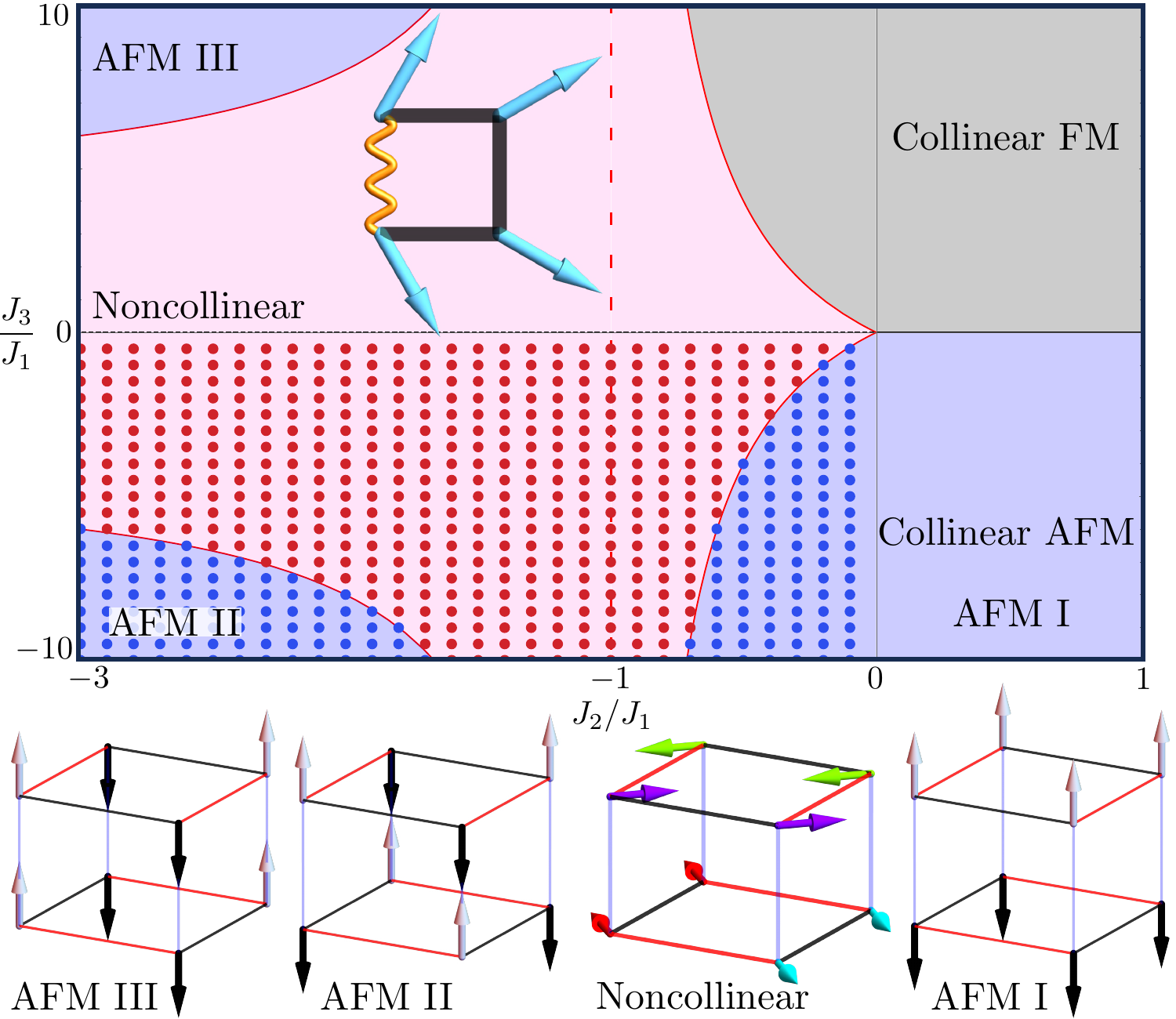}
  \caption{(Top) Ground-state phase diagram and (Bottom) the representative spin configurations in each phase. The red, black, and blue regions represent the noncollinear, collinear ferromagnet, and collinear antiferromagnet phases, respectively. The red (blue) dots mean the numerically obtained noncollinear (collinear) ground states. The red solid curve is the exact phase boundary between the noncollinear and collinear phases [Eq.~\eqref{eq:phaseboundary}]. The red dashed line indicates $J_2 = -J_1$. The diagram inside the figure schematically illustrates the geometric frustration which the quartet of the N\'eel fields suffers from. The yellow curve represents the antiferromagnetic coupling and the black solid lines represent the ferromagnetic coupling. The black, red, and blue lines in the bottom figure represent the exchange interactions with the coefficients $J_1$, $J_2$, and $J_3$, respectively.}
  \label{fig:2}
\end{figure}

\emph{Phase diagram.}| The phase diagram of model~\eqref{eq:ham} is governed by its zeroth-order Hamiltonian $H^{(0)}$
\begin{align}\label{eq:ham0}
  H^{(0)} &= \int \frac{d^2 x}{4a^2}  \left(\mathcal{E}_\mathbf{M} + \mathcal{E}_\mathbf{N} \right)\,,
  \\
  \mathcal{E}_\mathbf{M}&= - 4 ( J_1 + J_2) (\mathbf{M}^A_1 \cdot \mathbf{M}^A_2 + \mathbf{M}^B_1 \cdot \mathbf{M}^B_2) \nonumber
  \\
  & \quad - 2 J_3 (\mathbf{M}^A_1 \cdot \mathbf{M}^B_1 + \mathbf{M}^A_2 \cdot \mathbf{M}^B_2) \,,
  \\
  \mathcal{E}_{\mathbf{N}} &= -  4 ( J_1 - J_2) (\mathbf{N}^A_1 \cdot \mathbf{N}^A_2 + \mathbf{N}^B_1 \cdot \mathbf{N}^B_2) \nonumber
  \\ & \quad - 2 J_3 (\mathbf{N}^A_1 \cdot \mathbf{N}^B_1 - \mathbf{N}^A_2 \cdot \mathbf{N}^B_2) \,.
\end{align}
The energy per unit cell from the magnetization fields $\mathcal{E}_\mathbf{M}$ has a minimum $\mathcal{E}_c = - 8|J_1 + J_2| - 4 |J_3|$ when $\mathbf{M}^A_1 = \mathrm{sgn}(J_1+J_2)\mathbf{M}^A_2 = \mathrm{sgn}(J_3)\mathbf{M}^B_1 = \mathrm{sgn}(J_1+J_2)\mathrm{sgn}(J_3)\mathbf{M}^B_2$, which form a collinear order. Unlike the magnetization fields, the N\'eel fields suffer from the frustration and form a noncollinear but coplanar order when they minimizes $\mathcal{E}_\mathbf{N}$. At the minimum we can write the energy from the N\'eel fields as $\mathcal{E}_\mathbf{N} = - 8 ( J_1 - J_2) \cos (\theta_1 - \theta_2) - 2 J_3 (\cos 2\theta_1 - \cos 2\theta_2)$, where $\mathbf{N}^A_{1,2} = \mathcal{R} (\cos \theta_{1,2}, \sin \theta_{1,2}, 0)$ and $\mathbf{N}^B_{1,2} = \mathcal{R}(\cos \theta_{1,2}, -\sin \theta_{1,2}, 0)$ with a $\mathrm{SO}(3)$ order parameter $\mathcal{R}$. Furthermore, differentiating $\mathcal{E}_\mathbf{N}$ with respect to $\theta_1$ and $\theta_2$ gives the minimum $\mathcal{E}_{nc} = -4 \sqrt{4(J_1 - J_2)^2 + J_3^2}$ at 
\begin{align}
  \theta_1 = \frac{\pi}{2} - \theta_2\,,\quad  \tan 2 \theta_2 = \frac{2(J_1 - J_2)}{ -J_3}\,.
\end{align}
The phase boundary between collinear and noncollinear phases is determined by the condition $\mathcal{E}_c = \mathcal{E}_{nc}$, which gives the critical curve
\begin{align}\label{eq:phaseboundary}
  |J_3| = -\frac{4J_1J_2}{|J_1 + J_2|}\,, \quad (J_2 \leq 0)\,.
\end{align}
See Fig.~\ref{fig:2} for the phase diagram, we find the noncollinear, collinear ferromagnetic, and collinear antiferromagnetic phases, which are separated by the critical curve~\eqref{eq:phaseboundary}.
In the noncollinear phase, the magnetization fields $\mathbf{M}^{A,B}_{1,2}$ vanish and the quartet of the N\'eel fields forms a rigid body which breaks $\mathrm{SO}(3)$ symmetry completely. From the fully broken $\mathrm{SO}(3)$ symmetry, the effective field theory for the noncollinear phase is a $\mathrm{SO}(3)$ sigma model~\footnote{We refer to the $\mathcal{M}$ sigma model as a nonlinear sigma model whose order parameter manifold is $\mathcal{M}$. The model commonly denoted as the $\mathrm{O}(3)$ nonlinear sigma model corresponds to the $S^2$ sigma model.}. By contrast, in the collinear phases, the N\'eel fields vanish and the magnetization fields align collinearly. Due to the collinear order, the ground states have the unbroken $\mathrm{SO}(2)$ symmetry and the effective field theories are given by $\mathrm{SO}(3)/\mathrm{SO}(2) = S^2$ sigma models.
The collinear phases feature one ferromagnetic phase and three distinguishable antiferromagnetic phases: AFM~I, AFM~II, and AFM~III. In the AFM~I phase, the spins align ferromagnetically within each layer but align antiferromagnetically between the layers. In the AFM~II phase, all the spins have antiferromagnetic ordering. In the AFM~III phase, the spins within each layer have antiferromagnetic ordering, but the interlayer ordering is ferromagnetic. While the AFM~I phase has been considered in Ref.~\cite{Jin2024b}, the AFM~II and AFM~III phases have not been discussed before.
Note that, in the collinear antiferromagnetic phases, the ``N\'eel'' order is produced by the collinear alignment of the magnetization fields.

To verify our analytical predictions for the phase transition, we directly minimize the microscopic Hamiltonian~\eqref{eq:ham} by the Monte Carlo simulations~\cite{Hale2000, Amari2022} combined with a gradient-descent algorithm known as the arrested Newton flow~\cite{Speight2020, Gudnason2020, Gudnason2022c, Leask2022} (see the Supplemental Material~\cite{smalt} for details). Since the system is symmetric under the simultaneous sign change of $J_3$ and $\mathbf{S}^B$, we perform the numerical simulation in the region $J_3<0$. The resulting numerical data (see Fig.~\ref{fig:2}) are in excellent agreement with the exact phase boundary~\eqref{eq:phaseboundary}. This is our first main result: the noncollinear phase, the three distinct collinear antiferromagnetic phases, and the collinear ferromagnetic phase are found and the exact phase boundary between the noncollinear and the collinear phases is given by Eq.~\eqref{eq:phaseboundary}.

\emph{Field theory.}|To complete the formulation of the field theory for the noncollinear phase, we investigate the quadratic-derivative term $H^{(2)}$ and derive the kinetic term from the Wess-Zumino action.
Neglecting the magnetization fields, the quadratic-derivative term straightforwardly obtained by the gradient expansion is
\begin{equation}\label{eq:hamquad}
\begin{aligned}
  H^{(2)}_\mathrm{eff} &= -\int d^2x \frac{1}{2} \sum_{i=1,2} \mathrm{Tr}\left[ \partial_i \mathcal{R}^T \partial_i \mathcal{R} \Gamma_i \right]\,,
  \\
  \Gamma_1 &= -J_1 | n^A_2 \rangle \langle n^A_1 | + J_2 | n^B_2 \rangle \langle n^B_1 |\,,
  \\
  \Gamma_2 &= J_2 | n^A_2 \rangle \langle n^A_1 | - J_1 | n^B_2 \rangle \langle n^B_1 |\,.
\end{aligned}
\end{equation}
Here, the linear map $|v\rangle \langle w|$ is defined by $|v\rangle \langle w| q = (q\cdot w) v$ for arbitrary three-dimensional vectors $v$, $w$, and $q$. The $\mathrm{SO}(3)$ field $\mathcal{R}$ is defined by the relation
\begin{align}\label{eq:defR}
  \mathbf{N}^{A,B}_{1,2}(\mathbf{x},t) = \mathcal{R}(\mathbf{x},t) n^{A,B}_{1,2}\,,
\end{align}
with arbitrary unit coplanar vectors $n^{A,B}_{1,2}$ satisfying $n^{A,B}_1 \cdot n^{A,B}_2 = \cos (\theta_1 - \theta_2)$ and $ n^A_{1,2} \cdot n^B_{1,2} = \cos 2\theta_{1,2}$.

The kinetic term arises from the Wess-Zumino action~\cite{Wess1971, Witten1983, Fradkin2013, Sachdev2011, Han2017, Grensing2021}, defined by the topological term associated with $\pi_2(S^2)$
\begin{align}\label{eq:wz}
  S_\mathrm{WZ}[\mathbf{S}] = \int dt \int_0^1 du \, \mathbf{S} \cdot \left( \partial_t \mathbf{S} \times \partial_u \mathbf{S} \right)\,,
\end{align}
where $u$ is a coordinate on the artificial dimension. We compute the summation of the Wess-Zumino action over the all spins. After integrating out the magnetization fields, we obtain the effective action $S_{\mathrm{eff}} = S_\mathrm{kin} - \int dt\, H^{(2)}_\mathrm{eff}$, where the kinetic term in terms of the $\mathrm{SO}(3)$ fields $\mathcal{R}$ is given by
\begin{align}\label{eq:kinetic}
  S_\mathrm{kin} = &\int \frac{d^2x}{4a^2} dt \sum_{IJij}\bigg[C^{IJ}_{ij} \mathrm{Tr} \left[\partial_t \mathcal{R}^T \partial_t \mathcal{R} |n^I_i\rangle \langle n^J_j| \right] \nonumber
  \\
  &+ D^{IJ}_{ij} \left(\mathrm{Tr} \left[ \mathcal{R}^T \partial_t \mathcal{R} |n^I_i\rangle \langle n^J_j| \right]\right)^2 
  \\
  &+ F^{IJ}_{ij} \mathrm{Tr}\left[ \mathcal{R}^T\partial_t \mathcal{R} |n^I_i\rangle \langle \gamma_0|\right]\mathrm{Tr}\left[\mathcal{R}^T \partial_t \mathcal{R} |n^J_j\rangle \langle \gamma_0|\right] \bigg]\,, \nonumber
\end{align}
where $\gamma_0 = n^B_2 \times n^A_1$.
This is our second main result: the effective action consisting of Eqs.~\eqref{eq:hamquad} and \eqref{eq:kinetic} is the first example of the effective field theory for noncollinear altermagnets derived from a microscopic model.
See the End Matter and the Supplemental Material~\cite{smalt} for the derivation of Eq.\eqref{eq:kinetic} and the explicit expressions of the coefficients.

The effective action of the noncollinear phase is simplified in the weak-coupling limit $|J_3| \ll -J_2 \sim J_1$. In this limit, the angles for the coplanar order take the value $\theta_1 = \theta_2 =\pi/4$ and the N\'eel fields form an orthogonal \textit{spin frame}~\cite{Pradenas2024, Pradenas2024a, Tchernyshyov2024a, Wernert2025, Pradenas2025}. Note that, after taking the limit $J_3 \to 0$, the inter-layer coupling $J_3$ does not explicitly appear in the effective action. However, its effect remains as the constraint $n^A \cdot n^B = 0$, where $n^A=n^A_1=n^A_2$ and $n^B=n^B_1=n^B_2$. The effective action in the weak-coupling limit is given by
\begin{align}\label{eq:weaklimit}
  S_\mathrm{eff} \approx \int d^2x dt  \sum_\mu \frac{1}{2} \mathrm{Tr}\left[ \partial_\mu \mathcal{R}^T \partial_\mu \mathcal{R} \Gamma_\mu \right]\,,
\end{align}
where
\begin{align}
  \Gamma_t &= -\frac{1}{4a^2 J_2} \left( |n^A\rangle \langle n^A | + |n^B\rangle \langle n^B| \right)\,, \nonumber
  \\
  \Gamma_x &= -J_1 |n^A \rangle \langle n^A | + J_2 | n^B \rangle \langle n^B|\,,
  \\
  \Gamma_y &= J_2 |n^A \rangle \langle n^A | - J_1 | n^B \rangle \langle n^B|\,. \nonumber
\end{align}

\begin{figure}[t]
  \includegraphics[width=1\columnwidth]{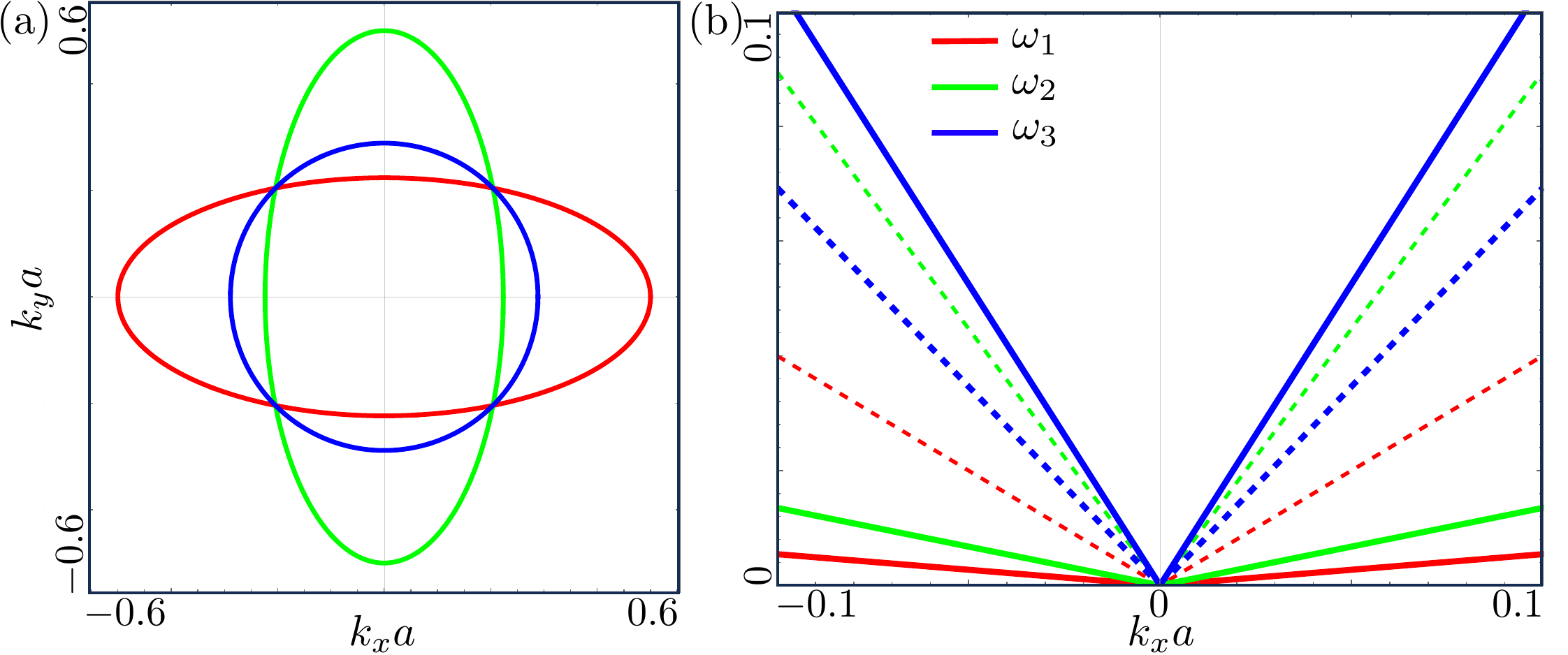}
  \caption{(a) Top view of the band structure of the $\mathrm{SO}(3)$ magnons. The red, green, and blue curves are the level sets $\omega_i (k_x, k_y) = 0.05$ for $i=1,2,3$, respectively. Here, $J_1=1, J_2=-0.2, J_3=0$. (b) Linear dispersion of the $\mathrm{SO}(3)$ magnons. The solid (dashed) line represents the dispersion relation with the parameters $J_1=1, J_2=-0.2$, and $J_3=-0.99\,(J_3 = 0)$.}
  \label{fig:band}
\end{figure}

\emph{Magnon.}|Magnons are elementary excitations of ordered spins. Unlike the collinear phases in which magnons reside on the tangent space of $S^2$, magnons in the noncollinear phase, dubbed $\mathrm{SO}(3)$ magnons, are defined on the tangent space of $\mathrm{SO}(3)$~\cite{Zarzuela2025a}. Since the order parameter manifold is three-dimensional, we naturally introduce three $\mathrm{SO}(3)$ magnon fields $\phi^a$ ($a=1,2,3$) as fluctuations on top of the ground state $\mathcal{R} \approx \mathds{1} + \phi^a L_a$, with $n^A_1 = (1,0,0)$, $n^A_2 = (\sin 2 \theta_2, - \cos 2\theta_2 , 0)$, $n^B_1=(-\cos 2\theta_2, - \sin 2 \theta_2, 0)$, and  $n^B_2 =(0,-1,0)$, where $L_a$ are the generators of $\mathrm{SO}(3)$. We find the effective action for the magnon fields from Eqs.~\eqref{eq:hamquad} and \eqref{eq:kinetic}, which is given by
\begin{align}\label{eq:sw}
  S_\mathrm{sw} &= \int d^2x dt \sum_{\mu ab}\xi_{\mu ab} \partial_\mu \phi^a \partial_\mu \phi^b \,,
\end{align}
where $\xi_t = \mathrm{diag}(\rho_1, \rho_1, \rho_3)$~\footnote{For brevity of analytical expressions, we have not considered the last term in Eq.~\eqref{eq:kinetic}, which is related to the orthogonality constraints. The contribution of this term simply enters as a correction $\rho_1 \to \rho_1 - \tilde{\rho}$. For a small $J_3$, this shift is negligible, while it attains its maximal magnitude at the phase boundary, where $\tilde{\rho} = \rho_1/4$. In contrast, for Fig.~\ref{fig:band} (b), we have taken this correction into account when plotting the bands at the phase boundary. The correction modifies the band slope and thereby the magnon velocity.}, and $\xi_x$ and $\xi_y$ are given by
\begin{align}
  \xi_x &= \frac{\sin 2 \theta_2 }{2}\begin{pmatrix}
    J_2 & -\frac{J_1 + J_2}{2\tan 2\theta_2} & 0 \\
    -\frac{J_1 + J_2}{2\tan 2\theta_2} & -J_1  & 0 \\
    0 & 0 & - (J_1 -J_2)
  \end{pmatrix}\,, \nonumber
  \\
  \xi_y &= \frac{\sin 2 \theta_2 }{2}\begin{pmatrix}
    -J_1 & \frac{J_1 + J_2}{2\tan 2\theta_2} & 0 \\
    \frac{J_1 + J_2}{2\tan 2\theta_2} & J_2  & 0 \\
    0 & 0 & - (J_1 -J_2)
  \end{pmatrix}\,,
\end{align}
with $\rho_1\! =\!(2\rho_\mathrm{diag} + \cos^2 (2\theta_2) \rho_\mathrm{inter} + \sin^2 (2\theta_2) \rho_\mathrm{intra})/(4a^2)$, and $\rho_3 =2(2\rho_\mathrm{diag} + \rho_\mathrm{inter} + \rho_\mathrm{intra} + \rho_\mathrm{ortho})/(4a^2)$. The coefficients $\rho_\mathrm{diag}$, $\rho_\mathrm{inter}$, $\rho_\mathrm{intra}$, and $\rho_\mathrm{ortho}$ are expressed in terms of $J_1$, $J_2$, and $J_3$, see the End Matter.
The magnon action~\eqref{eq:sw} is diagonalized as
\begin{align}
  S_\mathrm{sw} &= \int d^2x dt\bigg[ \rho_1(\partial_t \tilde{\phi}^1 )^2 - \lambda_1(\partial_x \tilde{\phi}^1)^2 - \lambda_2(\partial_y \tilde{\phi}^1)^2 \nonumber
  \\
  & +  \rho_1(\partial_t \tilde{\phi}^2 )^2 - \lambda_2(\partial_x \tilde{\phi}^2)^2 - \lambda_1(\partial_y \tilde{\phi}^2)^2 
  \\
  & + \rho_3 (\partial_t \phi^3 )^2 - \lambda_3(\partial_x \phi^3)^2 - \lambda_3(\partial_y \phi^3)^2 \bigg]\,, \nonumber
\end{align}
with the rotated fields $\tilde{\phi}^1 = \cos(\theta_2 - \frac{\pi}{4})\phi^1 - \sin(\theta_2 - \frac{\pi}{4})\phi^2$, $
\tilde{\phi}^2 = \sin(\theta_2 - \frac{\pi}{4})\phi^1 + \cos(\theta_2 - \frac{\pi}{4})\phi^2$, and positive coefficients $\lambda_1 = (-(J_1 + J_2) + (J_1 - J_2) \sin 2\theta_2)/4$, $\lambda_2 =(J_1 + J_2 + (J_1 - J_2) \sin 2\theta_2)/4$, and $\lambda_3 = (J_1 - J_2)\sin 2\theta_2 /2$. The diagonalized action gives the following linear dispersion relations for the $\mathrm{SO}(3)$ magnons:
\begin{gather}\label{eq:disp}
  \omega_1 = \sqrt{(\lambda_1 k_x^2 + \lambda_2 k_y^2)/\rho_1}\,, \quad  \omega_2 = \sqrt{(\lambda_2 k_x^2 + \lambda_1 k_y^2)/\rho_1} \,, \nonumber
  \\
  \omega_3 = \sqrt{\lambda_3 (k_x^2 +  k_y^2)/\rho_3}\,.
\end{gather}
Figures~\ref{fig:band}(a) and (b) respectively present top-down and side-view projections of the magnon dispersion. The top-down view clearly exhibits the characteristic $d$-wave anisotropy of altermagnetic systems~\cite{Smejkal2022, Smejkal2022a}. As the system approaches the phase boundary between the noncollinear phase and the AFM~I phase, the group velocities of the $\omega_1$ and $\omega_2$ modes vanish.
Moreover, the magnon velocities along the $x$-direction $c_i = d\omega_i/dk_x$ satisfy the relation $c_1^2 + c_2^2 = 2c_3^2$ in the weak-coupling limit, analogous to the identity presented in Ref.~\cite{Pradenas2024}.

\begin{figure}[t]
  \includegraphics[width=1\columnwidth]{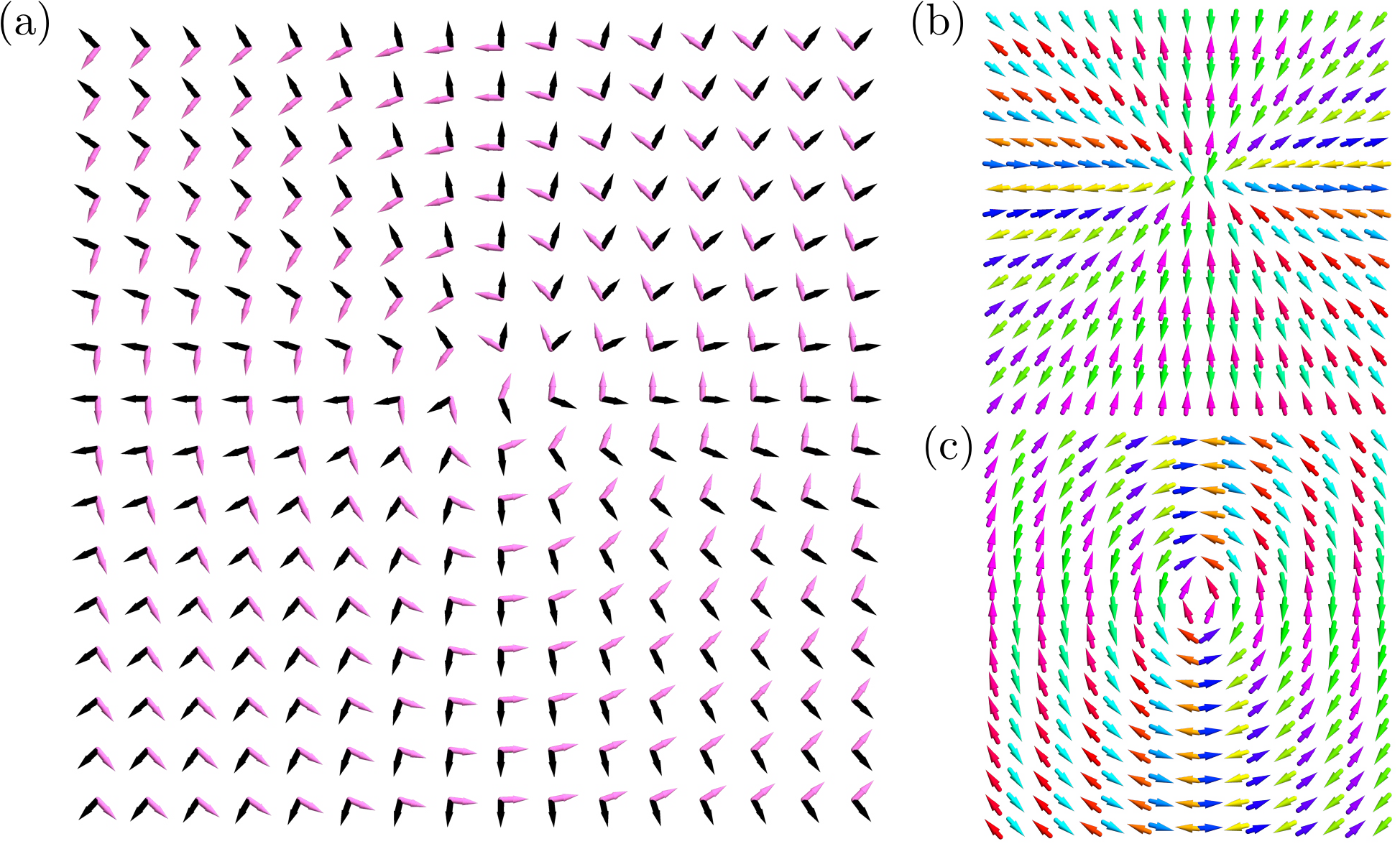}
  \caption{(a) Vortex configuration of the $\mathrm{SO}(3)$ order parameter. The black and pink arrows represent $\mathcal{R} n^A_1$ and $\mathcal{R} n^B_1$. Corresponding microscopic spin configurations in (b) layer $A$ and (c) layer $B$. Hue of the spins indicates the in-plane direction.}
  \label{fig:vortex}
\end{figure}

\emph{Vortex.}|Topological defects defined by the nontrivial homotopy $\pi_1(\mathrm{SO}(3))=\mathbb{Z}_2$, called $\mathbb{Z}_2$ vortices, appear in $\mathrm{SO}(3)$ sigma models. As can be inferred from the homotopy group, the $\mathbb{Z}_2$ vortex is its own antiparticle. This implies that, unlike vortices defined by $\pi_1(S^1)=\mathbb{Z}$ and commonly encountered in superfluids and superconductors, the net winding number cannot be increased more than one~\cite{Ochoa2018}.

We present a vortex configuration of the $\mathrm{SO}(3)$ field in Fig.~\ref{fig:vortex}(a). The solution is obtained by the arrested Newton flow from the initial parameterization $\mathcal{R}(x, y) = R_z (\varphi + \pi)$, $n^A_1 = n^A_2 = (1,0,0)$, and $n^B_1 = n^B_2 = (0,1,0)$, where $R_z$ is the rotation matrix around the $z$-axis and $\varphi$ is the polar angle in the $xy$-plane. To suppress unwanted finite-boundary effects, we consider a sufficiently large $2\times 256 \times 256$ system and exhibit the region in the vicinity of the vortex core. The parameters $J_1 = 1, J_2 = -0.5, J_3 = -0.1$ are used.

\emph{Discussion.}|In addition to the $\mathbb{Z}_2$ vortex, our $\mathrm{SO}(3)$ sigma model supports a variety of topological solitons through extensions of the model, such as incorporating the magnetocrystalline anisotropies or considering three-dimensional cases.
One of the most natural extensions includes the easy-plane anisotropy, which preserves the shape of the rigid body formed by the N\'eel fields but just projects the vacuum manifold onto $S^1$. Consequently, the topological defects associated with $\pi_1(S^1)=\mathbb{Z}$ are supported, and the easy-plane anisotropy prohibits multiple-winding vortices from smooth unwinding via $\mathrm{SO}(3)$ rotations.
The second possible extension considering the cubic or square anisotropy, a higher-order anisotropy along multiple easy-axes~\cite{Skomski2008, Voinescu2020, Lee2024}. When we take $x$- and $y$-axes as the easy-axes, the N\'eel fields may form an orthogonal frame aligned with the $x$- and $y$-axes in the ground states. This fact guarantees the discrete symmetry of the ground states and thereby domain walls associated with the discrete symmetry are supported. Domain walls in a kagome antiferromagnet with three easy-axes have been studied in Ref.~\cite{Ulloa2016}.

It is also worth considering the three-dimensional generalization and gauged models from the effective action~\eqref{eq:weaklimit}. The former includes stacking multiple layers, and the latter would be induced from the Dzyaloshinskii–Moriya interaction~\cite{Moriya1960}, similar to the case of collinear magnets~\cite{Schroers2019, Hill2021, Amari2023b, Amari2024h}. In the presence of the Dzyaloshinskii–Moriya interaction, one can anticipate a spatially modulated texture of $\mathrm{SO}(3)$ fields as a ground state analogous to the well-known magnetic chiral soliton lattices in collinear magnets~\cite{Togawa2012, Amari2024a, Amari2025a}. In three dimensions, the system has three-dimensional Skyrmions defined by $\pi_3(\mathrm{SO}(3))=\mathbb{Z}$. Three-dimensional skyrmions in condensed-matter systems have been studied in frustrated magnets~\cite{Zarzuela2019a, Naya2022, Zarzuela2025a}. However, the mechanisms governing their stabilization and the formation of skyrmion crystals remain poorly understood. Extending this model to a three-dimensional setting with Dzyaloshinskii–Moriya–type interactions could provide a promising platform for realizing stable $\mathrm{SO}(3)$ skyrmions and the corresponding crystal phases.

Our work enables the derivation of the hitherto unreported effective field theories for the noncollinear phase and collinear phases~\cite{smalt} as well as the effective field theory for the known collinear phase~\cite{Gomonay2024a, Jin2024b, Yershov2024b, Yershov2025}. These results provide a framework for advancing theoretical studies of noncollinear altermagnets.

\begin{acknowledgments}
  S.L. thanks Seunghun Lee for helpful discussions.
  This work was supported by the Brain Pool Plus Program through the National Research Foundation of Korea funded by the Ministry of Science and ICT (2020H1D3A2A03099291), the National Research Foundation of Korea(NRF) grant funded by the Korea government(MSIT) (2021R1C1C1006273), and the Basic Science Research Program through the National Research Foundation of Korea (NRF) funded by the Ministry of Education (2019R1A6A1A10073887).
\end{acknowledgments}

\bibliography{/Users/seungho/Documents/zot_library.bib}

\begin{appendix}
\renewcommand{\theequation}{A\arabic{equation}}
\setcounter{equation}{0}
\appendix
\section{End Matter}\label{endmatter}
The End Matter contains the derivation of the effective kinetic term~\eqref{eq:kinetic} from the Wess-Zumino action~\eqref{eq:wz}.
We compute the summation of the Wess-Zumino action over the all spins in layer $A$ as
\begin{equation}
\begin{aligned}
  &S^A_\mathrm{WZ} = \!\sum_{\mathbf{x} \in (2a \mathbb{Z})^2} \! \bigg[S_\mathrm{WZ}[\mathbf{S}_1^A(\mathbf{x})] + S_\mathrm{WZ}[\mathbf{S}_2^A(\mathbf{x} + a \hat{x})] 
  \\
  &+ S_\mathrm{WZ}[\mathbf{S}_3^A(\mathbf{x} + a \hat{x} + a \hat{y})] + S_\mathrm{WZ}[\mathbf{S}_4^A(\mathbf{x} + a \hat{y})]\bigg]\,.
\end{aligned}
\end{equation}
Neglecting the higher-order terms of the magnetization fields, the integration of the Wess-Zumino action gives
\begin{align}
  &S^A_\mathrm{WZ}  \approx \int \frac{d^2x}{4a^2} dt \bigg[  \mathbf{N}^A_1 \cdot \left( \partial_t \mathbf{N}^A_1 \times a (\partial_x + \partial_y) \mathbf{N}^A_1 \right) \nonumber
  \\
  &+ \mathbf{N}^A_2 \cdot \left( \partial_t \mathbf{N}^A_2 \times a (- \partial_x + \partial_y) \mathbf{N}^A_2 \right)
  \\
  & + 2 \mathbf{M}^A_1\cdot \left( \mathbf{N}^A_1  \times \partial_t \mathbf{N}^A_1  \right) + 2 \mathbf{M}^A_2\cdot \left( \mathbf{N}^A_2  \times \partial_t \mathbf{N}^A_2  \right) \bigg]\,. \nonumber
\end{align}
The first two terms are topological terms associated with $\pi_2(S^2)$, which do not contribute the equation of motion~\cite{Dombre1988, Fradkin2013, Sachdev2011, Grensing2021}. The same holds for layer $B$. Therefore, without the topological terms, the leading-order action can be written as
\begin{align}
  S & = \mathcal{S}_\mathrm{WZ} - \int dt \, H^{(0)} - \int dt\,  H^{(2)}_\mathrm{eff}\,,
  \\
  \mathcal{S}_\mathrm{WZ} &= S^A_\mathrm{WZ} + S^B_\mathrm{WZ}\,, \nonumber
  \\
  & \approx \int \frac{d^2x}{4a^2} dt \sum_{I=A,B}\sum_{i=1,2} 2\mathbf{M}^I_i\cdot \left( \mathbf{N}^I_i  \times \partial_t \mathbf{N}^I_i  \right)\,.
\end{align}
Since the N\'eel fields are dominant for describing the low-energy physics compared to the magnetization fields, we seek the effective field theory in terms of the N\'eel fields and the corresponding $\mathrm{SO}(3)$ order parameter.
The effective action obtained by integrating out the magnetization fields under the constraints $\mathbf{N}^{A,B}_{1,2} \cdot \mathbf{M}^{A,B}_{1,2} = 0$ is written as
\begin{align}\label{end:neelkin}
  &S_\mathrm{kin} = \int \frac{d^2x}{4a^2} dt \bigg[ \rho_\mathrm{diag}\sum_{I,\,i}|\partial_t \mathbf{N}^I_i|^2 
  \\
  &+ \rho_\mathrm{intra}\sum_I\left(\sin2\theta_2 \partial_t \mathbf{N}^I_1 \cdot \partial_t \mathbf{N}^I_2 + (\mathbf{N}^I_1 \cdot \partial_t \mathbf{N}^I_2)^2 \right) \nonumber
  \\
  &+ \rho_\mathrm{inter}\sum_i\left(\cos 2\theta_i \partial_t \mathbf{N}^A_i \cdot \partial_t \mathbf{N}^B_i + (\mathbf{N}^A_i \cdot \partial_t \mathbf{N}^B_i)^2 \right) \nonumber
  \\
  &+ \rho_\mathrm{ortho}\left(\left( \mathbf{N}^A_1 \cdot \partial_t \mathbf{N}^B_2 \right)^2 + \left( \mathbf{N}^A_2 \cdot \partial_t \mathbf{N}^B_1 \right)^2 \right) - \frac{1}{2} j^T A^{-1} j\bigg]\,, \nonumber
\end{align}
with
\begin{equation}\label{{end:rhos}}
\begin{aligned}
  \rho_\mathrm{diag} &= \frac{\mathcal{E}_{nc}J_1 J_2}{4\left[(4J_1 J_2)^2 - (J_1 + J_2)^2 J_3^2\right]}\,,
  \\
  \rho_\mathrm{intra} &= \frac{(J_1 + J_2)\left(J_3^2 -8 J_1 J_2   \right)}{2\left[(4J_1 J_2)^2 - (J_1 + J_2)^2 J_3^2\right]}\,,
  \\
  \rho_\mathrm{inter} &= \frac{\left(J_1^2 + J_2^2\right)J_3}{\left[(4J_1 J_2)^2 - (J_1 + J_2)^2 J_3^2\right]}  \,,
  \\
  \rho_\mathrm{ortho} &= -\frac{\mathcal{E}_{nc}\left(J_1 + J_2\right)J_3}{8\left[(4J_1 J_2)^2 - (J_1 + J_2)^2 J_3^2\right]}  \,.
\end{aligned}
\end{equation}
The last term in Eq.~\eqref{end:neelkin} stems from the constraints $\mathbf{N}^{A,B}_{1,2} \cdot \mathbf{M}^{A,B}_{1,2} = 0$, see the Supplemental Material~\cite{smalt} for details.
Furthermore, from the relation~\eqref{eq:defR}, we express the effective action in terms of the $\mathrm{SO}(3)$ field $\mathcal{R}$
\begin{align}\label{end:kin}
S_\mathrm{kin} = &\int \frac{d^2x}{4a^2} dt \sum_{IJij}\bigg[C^{IJ}_{ij} \mathrm{Tr} \left[\partial_t \mathcal{R}^T \partial_t \mathcal{R} |n^I_i\rangle \langle n^J_j| \right] \nonumber
\\
&+ D^{IJ}_{ij} \mathrm{Tr} \left[ \mathcal{R}^T \partial_t \mathcal{R} |n^I_i\rangle \langle n^J_j| \right]^2 
\\
&+ F^{IJ}_{ij} \mathrm{Tr}\left[ \mathcal{R}^T\partial_t \mathcal{R} |n^I_i\rangle \langle \gamma_0|\right]\mathrm{Tr}\left[\mathcal{R}^T \partial_t \mathcal{R} |n^J_j\rangle \langle \gamma_0|\right] \bigg]\,. \nonumber
\end{align}
where $\gamma_0 = n^B_2\times n^A_1$.
Here, the coefficients $C^{IJ}_{ij}$, $D^{IJ}_{ij}$, and $F^{IJ}_{ij}$, which are symmetric under the exchange of indices $(I \leftrightarrow J)$ and $(i \leftrightarrow j)$, are given by $C^{II}_{ii} = \rho_\mathrm{diag}$, $C^{AA}_{12} = C^{BB}_{12}=\rho_\mathrm{intra} \sin 2\theta_2/2$, $C^{AB}_{22} = -C^{AB}_{11} = \rho_\mathrm{inter}\cos 2\theta_2/2$, $C^{AB}_{12}=0$, $D^{II}_{ii}=0$, $D^{AA}_{12} = D^{BB}_{12} = \rho_\mathrm{intra}/2$, $D^{AB}_{22} = D^{AB}_{11} = \rho_\mathrm{inter}/2$, $D^{AB}_{12} = \rho_\mathrm{ortho}$, $F^{II}_{ii} = \mathcal{F}\mathcal{E}_{nc}^2 ( J_3^2 (4 (J_1^4 - 2 J_1^3 J_2 - 2 J_1^2 J_2^2 - 2 J_1 J_2^3 + J_2^4) + (J_1 + J_2)^2 J_3^2))/128  $, $F^{AB}_{12} = 0$,
$F^{AA}_{12} = F^{BB}_{12} = \mathcal{F} ((J_1^4 - J_2^4) J_3^2 (2 (J_1^2 - 4 J_1 J_2 + J_2^2) + J_3^2))$, and
$F^{AB}_{11} = - F^{AB}_{22} = \mathcal{F} (-4 J_1 J_2 (J_1^2 - J_2^2)^2 J_3^2 + (J_1^4 - J_1^3 J_2 - 2 J_1^2 J_2^2 - J_1 J_2^3 + J_2^4) J_3^4 +  (J_1 + J_2)^2 J_3^6/8)$, where $\mathcal{F} = 1/[\mathcal{E}_{nc} (J_1-J_2)^2 (J_3^2 - 8 J_1 J_2)[(4J_1 J_2)^2 -(J_1+J_2)^2 J_3^2]]$.
Likewise, the last term in Eq.~\eqref{end:kin} is originated from the constraints and is negligible with a small $J_3$.

\end{appendix}


\clearpage
\newpage

\onecolumngrid
\begin{center}
    \textbf{\large Supplemental Material: Effective Field Theory of a Noncollinear Altermagnet}\\
    \bigskip
    \text{Seungho Lee$^1$ and Se Kwon Kim$^1$}\\
    \medskip
    ${}^1$\textit{Department of Physics, Korea Advanced Institute of Science and Technology, Daejeon 34141, Republic of Korea} \\
\end{center}

\renewcommand{\thefigure}{S\arabic{figure}}

\renewcommand{\theequation}{\thesection.\,\arabic{equation}}

\renewcommand{\thetable}{S\Roman{table}}
\renewcommand{\thesection}{S\Roman{section}}
\renewcommand{\thesubsection}{\Alph{subsection}}
\renewcommand{\thesubsubsection}{\arabic{subsubsection}}

\setcounter{equation}{0}
\setcounter{figure}{0}
\setcounter{page}{1}

\setcounter{secnumdepth}{2}

\subsection*{Contents}

The Supplemental Material includes the following sections:
\begin{itemize}
  \item Section \ref{supp:ham}: derivation of the continuum Hamiltonian.
  \item Section \ref{supp:kinetic}: derivation of the kinetic term from the Wess-Zumino action.
  \item Section \ref{supp:col}: summary of the actions for the collinear phases.
  \item Section \ref{supp:numerical}: details of the numerical methods.
\end{itemize}

\bigskip

\section{Continuum Hamiltonian}\label{supp:ham}

In this section we derive the continuum Hamiltonian from the microscopic Hamiltonian~\eqref{eq:ham}.

\subsection{Zeroth-order term}

The Hamiltonians for layer $A$ and $B$ are written as
\begin{equation}
\begin{aligned}
  H_A = - \sum_{\mathbf{x}\in(2a \mathbb{Z})^2} &\bigg[ J_1 \mathbf{S}_1^A ( \mathbf{x}) \cdot \mathbf{S}_2^A(\mathbf{x} + a \hat{x}) + J_2 \mathbf{S}_2^A ( \mathbf{x} + a \hat{x}) \cdot \mathbf{S}_3^A(\mathbf{x} + a \hat{x} + a \hat{y})
  \\
  & + J_1 \mathbf{S}_3^A (\mathbf{x} +a \hat{x} +a \hat{y}) \cdot \mathbf{S}_4^A(\mathbf{x} + a\hat{y}) + J_2 \mathbf{S}_4^A(\mathbf{x} + a\hat{y}) \cdot \mathbf{S}_1^A (\mathbf{x})\bigg]\,,
  \\
  - \sum_{\mathbf{x}\in(2a \mathbb{Z})^2} &\bigg[ J_1 \mathbf{S}_2^A(\mathbf{x} + a\hat{x}) \cdot \mathbf{S}_1^A(\mathbf{x} + 2 a \hat{x}) + J_1 \mathbf{S}_3^A(\mathbf{x} + a\hat{x} + a\hat{y}) \cdot \mathbf{S}_4^A(\mathbf{x} + 2 a \hat{x} + a\hat{y})
  \\
  & + J_2 \mathbf{S}_3^A(\mathbf{x} + a\hat{x} + a\hat{y}) \cdot \mathbf{S}_2^A(\mathbf{x} + a \hat{x} +2a \hat{y}) + J_2 \mathbf{S}_4^A(\mathbf{x} + a\hat{y}) \cdot \mathbf{S}_1^A(\mathbf{x} + 2 a \hat{y})\bigg]\,,
\end{aligned}
\end{equation}
\begin{equation}
\begin{aligned}
  H_B = - \sum_{\mathbf{x}\in(2a \mathbb{Z})^2} &\bigg[ J_2 \mathbf{S}_1^A ( \mathbf{x}) \cdot \mathbf{S}_2^A(\mathbf{x} + a \hat{x}) + J_1 \mathbf{S}_2^A ( \mathbf{x} + a \hat{x}) \cdot \mathbf{S}_3^A(\mathbf{x} + a \hat{x} + a \hat{y})
  \\
  & + J_2 \mathbf{S}_3^A (\mathbf{x} +a \hat{x} +a \hat{y}) \cdot \mathbf{S}_4^A(\mathbf{x} + a\hat{y}) + J_1 \mathbf{S}_4^A(\mathbf{x} + a\hat{y}) \cdot \mathbf{S}_1^A (\mathbf{x})\bigg]\,,
  \\
  - \sum_{\mathbf{x}\in(2a \mathbb{Z})^2} &\bigg[ J_2 \mathbf{S}_2^A(\mathbf{x} + a\hat{x}) \cdot \mathbf{S}_1^A(\mathbf{x} + 2 a \hat{x}) + J_2 \mathbf{S}_3^A(\mathbf{x} + a\hat{x} + a\hat{y}) \cdot \mathbf{S}_4^A(\mathbf{x} + 2 a \hat{x} + a\hat{y})
  \\
  & + J_1 \mathbf{S}_3^A(\mathbf{x} + a\hat{x} + a\hat{y}) \cdot \mathbf{S}_2^A(\mathbf{x} + a \hat{x} +2a \hat{y}) + J_1 \mathbf{S}_4^A(\mathbf{x} + a\hat{y}) \cdot \mathbf{S}_1^A(\mathbf{x} + 2 a \hat{y})\bigg]\,.
\end{aligned}
\end{equation}
Note that the first summations are the intra-unit-cell interaction and the second summations are the inter-unit-cell interaction. To obtain a field theory, similar to the Haldane mapping, we introduce the N\'eel order parameters and the magnetization fields as
\begin{equation}
  \begin{aligned}
  \mathbf{S}_{1,3}^A = \mp \mathbf{N}_1^A + \mathbf{M}_1^A\,, \quad \mathbf{S}_{2,4}^A = \mp \mathbf{N}_2^A + \mathbf{M}_2^A\,,
  \\
  \mathbf{S}_{1,3}^B = \mp \mathbf{N}_1^B + \mathbf{M}_1^B\,, \quad \mathbf{S}_{2,4}^B = \pm \mathbf{N}_2^B + \mathbf{M}_2^B\,,
  \end{aligned}
\end{equation}
with the constraints $\mathbf{N}^{A,B}_{1,2} \cdot \mathbf{M}^{A,B}_{1,2} = 0$.
For the smoothness of the N\'eel order parameters, $\mathbf{S}^B_{2,4}$ and $\mathbf{N}^B_2$ are defined with an opposite sign to others. In the noncollinear phase, the N\'eel order parameters are dominant and the magnetization fields act as fluctuation modes.

Applying the gradient expansion, we collect up to the quadratic derivative terms
\begin{align}
  H = H^{(0)} + H^{(1)} + H^{(2)} + \cdots\,.
\end{align}
The zeroth-order terms of $H_A$ and $H_B$ are given by
\begin{align}
  H_A^{(0)} = - \sum_{\mathbf{x}\in(2a \mathbb{Z})^2} &2\bigg[ J_1 \mathbf{S}_1^A ( \mathbf{x}) \cdot \mathbf{S}_2^A(\mathbf{x} ) + J_2 \mathbf{S}_2^A ( \mathbf{x} ) \cdot \mathbf{S}_3^A(\mathbf{x} ) + J_1 \mathbf{S}_3^A (\mathbf{x} ) \cdot \mathbf{S}_4^A(\mathbf{x}) + J_2 \mathbf{S}_4^A(\mathbf{x}) \cdot \mathbf{S}_1^A (\mathbf{x})\bigg] \nonumber
  \\
  = - \sum_{\mathbf{x} \in (2a \mathbb{Z})^2} & 4 \bigg[ (J_1 - J_2) \mathbf{N}_1^A (\mathbf{x}) \cdot \mathbf{N}_2^A(\mathbf{x}) + (J_1 + J_2)\mathbf{M}_1^A(\mathbf{x}) \cdot \mathbf{M}_2^A(\mathbf{x}) \bigg]\,,
  \\
  H_B^{(0)} = - \sum_{\mathbf{x}\in(2a \mathbb{Z})^2} &2\bigg[ J_2 \mathbf{S}_1^B ( \mathbf{x}) \cdot \mathbf{S}_2^B(\mathbf{x} ) + J_1 \mathbf{S}_2^B ( \mathbf{x} ) \cdot \mathbf{S}_3^B(\mathbf{x} ) + J_2 \mathbf{S}_3^B (\mathbf{x} ) \cdot \mathbf{S}_4^B(\mathbf{x}) + J_1 \mathbf{S}_4^B(\mathbf{x}) \cdot \mathbf{S}_1^B (\mathbf{x})\bigg]\nonumber
  \\
  = - \sum_{\mathbf{x} \in (2a \mathbb{Z})^2} & 4 \bigg[ (J_1 - J_2) \mathbf{N}_1^B (\mathbf{x}) \cdot \mathbf{N}_2^B(\mathbf{x}) + (J_1 + J_2)\mathbf{M}_1^B(\mathbf{x}) \cdot \mathbf{M}_2^B(\mathbf{x}) \bigg]\,.
\end{align}
The inter-layer interaction $H_\mathrm{int}$ which endows the system with the frustration is written as
\begin{equation}
\begin{aligned}
  H_\mathrm{int} = - \sum_{\mathbf{x} \in ( 2a \mathbb{Z})^2} &J_3\bigg[  \mathbf{S}^A_1(\mathbf{x}) \cdot \mathbf{S}^B_1 (\mathbf{x}) + \mathbf{S}^A_2(\mathbf{x} + a \hat{x}) \cdot \mathbf{S}^B_2 (\mathbf{x} + a \hat{x}) \\
  & + \mathbf{S}^A_3(\mathbf{x} + a \hat{x} + a \hat{y}) \cdot \mathbf{S}^B_3 (\mathbf{x} + a \hat{x} + a \hat{y}) + \mathbf{S}^A_4(\mathbf{x} + a \hat{y}) \cdot \mathbf{S}^B_4 (\mathbf{x} + a \hat{y})\bigg]
  \\
  = - \sum_{\mathbf{x} \in ( 2a \mathbb{Z})^2} &J_3\bigg[  \left( -\mathbf{N}_1^A(\mathbf{x}) + \mathbf{M}_1^A(\mathbf{x})\right) \cdot \left( -\mathbf{N}_1^B(\mathbf{x}) + \mathbf{M}_1^B(\mathbf{x})\right)
  \\
  & +  \left( -\mathbf{N}_2^A(\mathbf{x} + a\hat{x}) + \mathbf{M}_2^A(\mathbf{x} + a\hat{x})\right) \cdot \left( \mathbf{N}_2^B(\mathbf{x} + a\hat{x}) + \mathbf{M}_2^B(\mathbf{x} + a\hat{x})\right)
  \\
  & +  \left( \mathbf{N}_1^A(\mathbf{x} + a\hat{x} + a\hat{y}) + \mathbf{M}_1^A(\mathbf{x} + a\hat{x} + a\hat{y})\right) \cdot \left( \mathbf{N}_1^B(\mathbf{x} + a\hat{x} + a\hat{y}) + \mathbf{M}_1^B(\mathbf{x} + a\hat{x} + a\hat{y})\right)
  \\
  & +  \left( \mathbf{N}_2^A(\mathbf{x} + a\hat{y}) + \mathbf{M}_2^A(\mathbf{x} + a\hat{y})\right) \cdot \left( -\mathbf{N}_2^B(\mathbf{x} + a\hat{y}) + \mathbf{M}_2^B(\mathbf{x} + a\hat{y})\right) \bigg]
  \\
  = - \int \frac{d^2x}{4a^2} & J_3  \left[ 2\left( \mathbf{N}_1^A \cdot \mathbf{N}_1^B - \mathbf{N}_2^A \cdot \mathbf{N}_2^B + \mathbf{M}_1^A \cdot \mathbf{M}_1^B + \mathbf{M}_2^A \cdot \mathbf{M}_2^B \right) \right]\,.
\end{aligned}
\end{equation}
Note that the inter-layer interaction only contains zeroth order derivative terms $H_\mathrm{int}= H_\mathrm{int}^{(0)}$. Therefore the zeroth order of the total Hamiltonian is
\begin{equation}
\begin{aligned}
  H^{(0)} = - \int \frac{d^2 x}{4a^2}  &\bigg[ 4(J_1 - J_2) (\mathbf{N}_1^A \cdot \mathbf{N}_2^A + \mathbf{N}_1^B \cdot \mathbf{N}_2^B) + 4(J_1 + J_2)(\mathbf{M}_1^A \cdot \mathbf{M}_2^A + \mathbf{M}_1^B \cdot \mathbf{M}_2^B)
  \\
  & + 2J_3\left( \mathbf{N}_1^A \cdot \mathbf{N}_1^B - \mathbf{N}_2^A \cdot \mathbf{N}_2^B + \mathbf{M}_1^A \cdot \mathbf{M}_1^B + \mathbf{M}_2^A \cdot \mathbf{M}_2^B \right) \bigg]\,.
\end{aligned}
\end{equation}
Due to the constraints $|\mathbf{N}^{A,B}_{1,2}|^2 + |\mathbf{M}^{A,B}_{1,2}|^2 = 1$, following two energy densities are not simultaneously minimized and the competition between them determines the phase of the ground states.
\begin{align}
  \mathcal{E}_{\mathbf{M}}&= - \left( 4 ( J_1 + J_2) (\mathbf{M}^A_1 \cdot \mathbf{M}^A_2 + \mathbf{M}^B_1 \cdot \mathbf{M}^B_2) + 2 J_3 (\mathbf{M}^A_1 \cdot \mathbf{M}^B_1 + \mathbf{M}^A_2 \cdot \mathbf{M}^B_2) \right)\,, \label{eq:em}
  \\
  \mathcal{E}_{\mathbf{N}} &= - \left( 4 ( J_1 - J_2) (\mathbf{N}^A_1 \cdot \mathbf{N}^A_2 + \mathbf{N}^B_1 \cdot \mathbf{N}^B_2) + 2 J_3 (\mathbf{N}^A_1 \cdot \mathbf{N}^B_1 - \mathbf{N}^A_2 \cdot \mathbf{N}^B_2) \right)\,. \label{eq:en}
\end{align}
In the collinear phases Eq.~\eqref{eq:em} is minimized from the collinear order of the magnetization fields
\begin{equation}
  \mathbf{M}^A_1 = \mathrm{sgn}(J_1+J_2)\mathbf{M}^A_2 = \mathrm{sgn}(J_3)\mathbf{M}^B_1 = \mathrm{sgn}(J_1+J_2)\mathrm{sgn}(J_3)\mathbf{M}^B_2
\end{equation}
and has a minimum
\begin{equation}
  \mathcal{E}_c = - 8|J_1 + J_2| - 4 |J_3|\,.
\end{equation}

\subsection{Linear derivative term}

Now let us consider the linear derivative term $H^{(1)}$ of the continuum Hamiltonian. The linear term generally vanishes in all phases, since
\begin{equation}
\begin{aligned}
  H_A^{(1)} = \int \frac{d^2x}{4a^2} &\bigg[ J_1 \mathbf{S}^A_1 \cdot a \partial_x \mathbf{S}^A_2 + J_2 a \partial_x \mathbf{S}^A_2 \cdot \mathbf{S}^A_3 + J_2 \mathbf{S}^A_2 \cdot ( a \partial_x + a \partial_y) \mathbf{S}^A_3
  \\
  &+ J_1 (a \partial_x + a \partial_y) \mathbf{S}^A_3 \cdot \mathbf{S}^A_4 + J_1 \mathbf{S}^A_3 \cdot a \partial_y \mathbf{S}^A_4 + J_2 a \partial_y \mathbf{S}^A_4 \cdot \mathbf{S}^A_1 
  \\
  &+ J_1 a \partial_x \mathbf{S}^A_2 \cdot \mathbf{S}^A_1 + J_1 \mathbf{S}^A_2 \cdot 2a \partial_x \mathbf{S}^A_1 + J_1 (a \partial_x + a \partial_y) \mathbf{S}^A_3 \cdot \mathbf{S}^A_4 + J_1 \mathbf{S}^A_3 \cdot (2a \partial_x + a \partial_y) \mathbf{S}^A_4
  \\
  &+ J_2 (a \partial_x + a \partial_y) \mathbf{S}^A_3 \cdot \mathbf{S}^A_2 + J_2 \mathbf{S}^A_3 \cdot (a \partial_x + 2a \partial_y) \mathbf{S}^A_2 + J_2 a \partial_y \mathbf{S}^A_4 \cdot \mathbf{S}^A_1 + J_2 \mathbf{S}^A_4 \cdot 2a\partial_y \mathbf{S}^A_1 \bigg]
  \\
  = \int \frac{d^2x}{4a^2} &\big[ J_1 \left(\mathbf{S}_1 \cdot a \partial_x \mathbf{S}_2 + a \partial_x \mathbf{S}_2 \cdot \mathbf{S}_1 + \mathbf{S}_2 \cdot 2 a \partial_x \mathbf{S}_1  \right)
  \\
  & + J_2 \left( a \partial_x \mathbf{S}_2 \cdot \mathbf{S}_3 + \mathbf{S}_2 \cdot (a \partial_x + a \partial_y) \mathbf{S}_3 + (a \partial_x + a \partial_y) \mathbf{S}_3 \cdot \mathbf{S}_2 + \mathbf{S}_3 \cdot (a \partial_x + 2 a \partial_y) \mathbf{S}_2\right)
  \\
  & + J_1 \left( (a \partial_x + a \partial_y) \mathbf{S}_3 \cdot \mathbf{S}_4  + \mathbf{S}_3 \cdot a \partial_y \mathbf{S}_4 + (a \partial_x + a \partial_y) \mathbf{S}_4 + \mathbf{S}_3 \cdot (2a \partial_x + \partial_y) \mathbf{S}_4 \right)
  \\
  & + J_2 \left( a \partial_y \mathbf{S}_4 \cdot \mathbf{S}_1 + a \partial_y \mathbf{S}_4 \cdot \mathbf{S}_1 + \mathbf{S}_4 \cdot 2a\partial_y  \mathbf{S}_1\right)\big] = 0\,,
\end{aligned}
\end{equation}
and similarly, $H_B^{(1)}=0$. 

\subsection{Quadratic derivative term}
The quadratic derivative terms from the Hamiltonian $H_A$ and $H_B$ are given by
\begin{align}
  H^{(2)}_A &= \int \frac{d^2x}{4}\bigg[ J_1 \partial_x\mathbf{S}^A_1 \cdot \partial_x\mathbf{S}^A_2 + J_2 \partial_y\mathbf{S}^A_2 \cdot \partial_y\mathbf{S}^A_3 + J_1 \partial_x\mathbf{S}^A_3 \cdot \partial_x\mathbf{S}^A_4 + J_2 \partial_y\mathbf{S}^A_4 \cdot \partial_y\mathbf{S}^A_1\bigg]\,,
  \\
  H^{(2)}_B &= \int \frac{d^2x}{4}\bigg[ J_2 \partial_x\mathbf{S}^B_1 \cdot \partial_x\mathbf{S}^B_2 + J_1 \partial_y\mathbf{S}^B_2 \cdot \partial_y\mathbf{S}^B_3 + J_2 \partial_x\mathbf{S}^B_3 \cdot \partial_x\mathbf{S}^B_4 + J_1 \partial_y\mathbf{S}^B_4 \cdot \partial_y\mathbf{S}^B_1\bigg]\,.
\end{align}
In terms of the N\'eel order fields and the magnetization fields, the quadratic Hamiltonian $H^{(2)} = H^{(2)}_A + H^{(2)}_B$ is written as
\begin{equation}
\begin{aligned}
  H^{(2)} = \int \frac{d^2x}{2} &\bigg[ J_1 \partial_x \mathbf{N}^A_1 \cdot \partial_x \mathbf{N}^A_2  - J_2 \partial_y \mathbf{N}^A_1 \cdot \partial_y \mathbf{N}^A_2 + J_1\partial_x \mathbf{M}^A_1 \cdot \partial_x\mathbf{M}^A_2 + J_2\partial_y \mathbf{M}^A_1 \cdot \partial_y\mathbf{M}^A_2
  \\
  & - J_2 \partial_x \mathbf{N}^B_1 \cdot \partial_x \mathbf{N}^B_2  + J_1 \partial_y \mathbf{N}^B_1 \cdot \partial_y \mathbf{N}^B_2 + J_2\partial_x \mathbf{M}^B_1 \cdot \partial_x\mathbf{M}^B_2 + J_1\partial_y \mathbf{M}^B_1 \cdot \partial_y\mathbf{M}^B_2\bigg]\,.
\end{aligned}
\end{equation}

In the noncollinear phase, neglecting the derivatives of the magnetization fields, we have the effective Hamiltonian
\begin{equation}
  H^{(2)}_\mathrm{eff} = \int \frac{d^2x}{2} \bigg[ J_1 \partial_x \mathbf{N}^A_1 \cdot \partial_x \mathbf{N}^A_2  - J_2 \partial_y \mathbf{N}^A_1 \cdot \partial_y \mathbf{N}^A_2 - J_2 \partial_x \mathbf{N}^B_1 \cdot \partial_x \mathbf{N}^B_2  + J_1 \partial_y \mathbf{N}^B_1 \cdot \partial_y \mathbf{N}^B_2\bigg]\,.
\end{equation}
This Hamiltonian can be expressed as a $\mathrm{SO}(3)$ sigma model
\begin{equation}
  \begin{aligned}
    H^{(2)}_\mathrm{eff} &= -\int d^2x \frac{1}{2} \sum_{i=1,2} \mathrm{Tr}\left[ \partial_i \mathcal{R}^T \partial_i \mathcal{R} \Gamma_i \right]\,,
    \\
    \Gamma_1 &= -J_1 | n^A_2 \rangle \langle n^A_1 | + J_2 | n^B_2 \rangle \langle n^B_1 |\,,
    \\
    \Gamma_2 &= J_2 | n^A_2 \rangle \langle n^A_1 | - J_1 | n^B_2 \rangle \langle n^B_1 |\,.
  \end{aligned}
\end{equation}
The Hamiltonian becomes isotropic at $J_2 = - J_1$.

In the collinear phases, the N\'eel fields are negligible and the quadratic Hamiltonian can be written as
\begin{equation}
  H^{(2)} \approx \int \frac{d^2x}{2} \bigg[ J_1 \partial_x \mathbf{M}^A_1 \cdot \partial_x \mathbf{M}^A_2 + J_2\partial_y \mathbf{M}^A_1 \cdot \partial_y\mathbf{M}^A_2 + J_2\partial_x \mathbf{M}^B_1 \cdot \partial_x\mathbf{M}^B_2 + J_1\partial_y \mathbf{M}^B_1 \cdot \partial_y\mathbf{M}^B_2\bigg]\,.
\end{equation}
In the collinear ferromagnetic phase,
\begin{equation}
  H^{(2)}_\mathrm{FM} = \int \frac{d^2x}{2} \left[ J_1 |\partial_x \mathbf{M}^A|^2 + J_2 |\partial_y \mathbf{M}^A|^2 + J_2 |\partial_x \mathbf{M}^B|^2 + J_1 |\partial_y \mathbf{M}^B|^2\right]\,.
\end{equation}
where
\begin{equation}\label{eq:fm}
  \mathbf{M}^A = \mathbf{M}^A_1 = \mathbf{M}^A_2\,, \quad \mathbf{M}^B = \mathbf{M}^B_1 = \mathbf{M}^B_2\,.
\end{equation}
In the AFM~I phase, to capture the antiferromagnetic ordering between the two layers, we introduce the following N\'eel order parameter and fluctuation
\begin{align}\label{eq:afm1}
  \mathbf{M}^A_1 = \mathbf{M}^A_2 = \mathbf{n} + \mathbf{m}\,, \quad \mathbf{M}^B_1 = \mathbf{M}^B_2 = -\mathbf{n} + \mathbf{m}\,.
\end{align}
With these variables, the quadratic Hamiltonian can be expressed as
\begin{equation}
\begin{aligned}
  H^{(2)}_\mathrm{AFMI} &= \int d^2x \frac{1}{2}\bigg[ (J_1+J_2) \sum_i\left(\left|\partial_i \mathbf{n}\right|^2 + \left|\partial_i \mathbf{m}\right|^2\right) + 2(J_1 - J_2) \left(\partial_x \mathbf{n} \cdot \partial_x \mathbf{m} - \partial_y \mathbf{n} \cdot \partial_y \mathbf{m}\right)\bigg]
  \\
  &\approx  \int d^2x \frac{1}{2}\bigg[ (J_1+J_2) \sum_i\left|\partial_i \mathbf{n}\right|^2  + 2(J_1 - J_2) \left(\partial_x \mathbf{n} \cdot \partial_x \mathbf{m} - \partial_y \mathbf{n} \cdot \partial_y \mathbf{m}\right)\bigg]
  \\
  &=\int d^2x \bigg[ \frac{(J_1+J_2)}{2} \sum_i\left|\partial_i \mathbf{n}\right|^2  - (J_1 - J_2) \,\mathbf{m}\cdot \left(\partial_x^2 \mathbf{n}    - \partial_y^2 \mathbf{n}   \right)\bigg]\,.
\end{aligned}
\end{equation}
Similarly, in the AFM~II phase,
\begin{equation}
\begin{aligned}
  H^{(2)}_\mathrm{AFMII} &\approx \int d^2x \frac{-(J_1+J_2) }{2}\sum_i\left(\left|\partial_i \mathbf{n}\right|^2 + 2 \partial_i \mathbf{n} \cdot \partial_i \mathbf{m} \right)
  \\
  &= \int d^2x \frac{-(J_1+J_2) }{2}\sum_i\left(\left|\partial_i \mathbf{n}\right|^2 - 2 \partial_i^2 \mathbf{n} \cdot  \mathbf{m} \right)\,.
\end{aligned}
\end{equation}
where
\begin{align}\label{eq:afm2}
  \mathbf{M}^A_1 = \mathbf{M}^B_2 = \mathbf{n} + \mathbf{m}\,, \quad \mathbf{M}^A_2 = \mathbf{M}^B_1 = -\mathbf{n} + \mathbf{m}\,.
\end{align}
Similarly, in the AFM~III phase,
\begin{equation}
\begin{aligned}
  H^{(2)}_\mathrm{AFMIII} &\approx \int d^2x \frac{-(J_1+J_2) }{2}\sum_i\left(\left|\partial_i \mathbf{n}\right|^2 + 2 \partial_i \mathbf{n} \cdot \partial_i \mathbf{m} \right)
  \\
  &= \int d^2x \frac{-(J_1+J_2) }{2}\sum_i\left(\left|\partial_i \mathbf{n}\right|^2 - 2 \partial_i^2 \mathbf{n} \cdot  \mathbf{m} \right)\,.
\end{aligned}
\end{equation}
where
\begin{align}\label{eq:afm3}
  \mathbf{M}^A_1 = \mathbf{M}^B_1 = \mathbf{n} + \mathbf{m}\,, \quad \mathbf{M}^A_2 = \mathbf{M}^B_2 = -\mathbf{n} + \mathbf{m}\,.
\end{align}

\section{Kinetic term}\label{supp:kinetic}
In this section we provide the details of the derivation of the kinetic terms for the noncollinear and collinear phases. We begin with the Wess-Zumino action defined by
\begin{align}
  S_\mathrm{WZ}[\mathbf{S}] = \int dt \int_0^1 du \, \mathbf{S} \cdot \left( \partial_t \mathbf{S} \times \partial_u \mathbf{S} \right)\,,
\end{align}
and integrate out fluctuation modes to obtain the effective actions.

\subsection{Noncollinear phase}
We first consider the noncollinear phase. For layer $A$, we compute
\begin{align}
  S^A_\mathrm{WZ} \equiv \sum_{\mathbf{x} \in (2a \mathbb{Z})^2} \left[S_\mathrm{WZ}[\mathbf{S}_1^A(\mathbf{x})] + S_\mathrm{WZ}[\mathbf{S}_2^A(\mathbf{x} + a \hat{x})] + S_\mathrm{WZ}[\mathbf{S}_3^A(\mathbf{x} + a \hat{x} + a \hat{y})] + S_\mathrm{WZ}[\mathbf{S}_4^A(\mathbf{x} + a \hat{y})]\right]\,.
\end{align}
We can expand the Wess-Zumino term as
\begin{equation}\label{eq:Swz}
\begin{aligned}
  S_\mathrm{WZ}[\mathbf{S}^A_1] &= S_\mathrm{WZ}[- \mathbf{N}^A_1  + \mathbf{M}^A_1]
  \\
  &= \int dt \int_0^1 du \, \left( -\mathbf{N}^A_1 + \mathbf{M}^A_1\right) \cdot \left( \partial_t \left(-\mathbf{N}^A_1 + \mathbf{M}^A_1\right) \times \partial_u \left(-\mathbf{N}^A_1 + \mathbf{M}^A_1\right)\right)
  \\
  & = \int dt \int_0^1 du \big[ - \mathbf{N}^A_1 \cdot \left( \partial_t \mathbf{N}^A_1 \times \partial_u \mathbf{N}^A_1\right) - \mathbf{N}^A_1 \cdot \left( \partial_t \mathbf{M}^A_1 \times \partial_u \mathbf{M}^A_1\right)
  \\
  & \quad +  \mathbf{N}^A_1 \cdot \left( \partial_t \mathbf{N}^A_1 \times \partial_u \mathbf{M}^A_1\right) + \mathbf{N}^A_1 \cdot \left( \partial_t \mathbf{M}^A_1 \times \partial_u \mathbf{N}^A_1\right)
  \\
  & \quad + \mathbf{M}^A_1 \cdot \left( \partial_t \mathbf{N}^A_1 \times \partial_u \mathbf{N}^A_1\right) + \mathbf{M}^A_1 \cdot \left( \partial_t \mathbf{M}^A_1 \times \partial_u \mathbf{M}^A_1\right)
  \\
  & \quad + \mathbf{M}^A_1 \cdot \left( \partial_t \mathbf{N}^A_1 \times \partial_u \mathbf{M}^A_1\right) + \mathbf{M}^A_1 \cdot \left( \partial_t \mathbf{M}^A_1 \times \partial_u \mathbf{N}^A_1\right) \big]\,.
\end{aligned}
\end{equation}
Since $|\mathbf{M}^A_{1,2}| \ll 1$, we neglect the second and third order of $\mathbf{M}^A_{1,2}$. Then the Wess-Zumino term becomes
\begin{equation}
\begin{aligned}
  S_\mathrm{WZ}[\mathbf{S}^A_1(\mathbf{x})]
  &= S_\mathrm{WZ}[- \mathbf{N}^A_1 (\mathbf{x}) + \mathbf{M}^A_1(\mathbf{x})]
  \\
  &\approx \int dt \int_0^1 du \big[ - \mathbf{N}^A_1(\mathbf{x}) \cdot \left( \partial_t \mathbf{N}^A_1(\mathbf{x}) \times \partial_u \mathbf{N}^A_1(\mathbf{x})\right)
  \\
  & + \mathbf{N}^A_1(\mathbf{x}) \cdot \left( \partial_t \mathbf{N}^A_1(\mathbf{x}) \times \partial_u \mathbf{M}^A_1(\mathbf{x})\right) + \mathbf{N}^A_1(\mathbf{x}) \cdot \left( \partial_t \mathbf{M}^A_1(\mathbf{x}) \times \partial_u \mathbf{N}^A_1(\mathbf{x})\right)]
  \\
  & = - \int dt \int_0^1 du \, \mathbf{N}^A_1(\mathbf{x}) \cdot \left( \partial_t \mathbf{N}^A_1(\mathbf{x}) \times \partial_u \mathbf{N}^A_1(\mathbf{x})\right)
  \\
  & \quad + \int dt \int_0^1 du \left[ \partial_u \left( \mathbf{N}^A_1(\mathbf{x}) \cdot ( \partial_t \mathbf{N}^A_1(\mathbf{x}) \times \mathbf{M}^A_1(\mathbf{x})) \right) +  \partial_t \left( \mathbf{N}^A_1(\mathbf{x}) \cdot (  \mathbf{M}^A_1(\mathbf{x}) \times \partial_u \mathbf{N}^A_1(\mathbf{x})) \right) \right]
  \\
  & = - \int dt \int_0^1 du \, \mathbf{N}^A_1(\mathbf{x}) \cdot \left( \partial_t \mathbf{N}^A_1(\mathbf{x}) \times \partial_u \mathbf{N}^A_1(\mathbf{x})\right) + \int dt \, \mathbf{M}^A_1(\mathbf{x})\cdot \left( \mathbf{N}^A_1 (\mathbf{x}) \times \partial_t \mathbf{N}^A_1 (\mathbf{x}) \right)
  \\
  & = -S_\mathrm{WZ}[\mathbf{N}^A_1(\mathbf{x})] + \int dt \, \mathbf{M}^A_1(\mathbf{x})\cdot \left( \mathbf{N}^A_1 (\mathbf{x}) \times \partial_t \mathbf{N}^A_1 (\mathbf{x}) \right)\,.
\end{aligned}
\end{equation}
For $\mathbf{S}^A_3$, we have
\begin{equation}
\begin{aligned}
  S_\mathrm{WZ}&[\mathbf{S}^A_3(\mathbf{x} + a \hat{x} + a \hat{y})] = S_\mathrm{WZ}[ \mathbf{N}^A_1 (\mathbf{x} + a \hat{x} + a \hat{y}) + \mathbf{M}^A_1(\mathbf{x} + a \hat{x} + a \hat{y})]
  \\
  &\approx S_\mathrm{WZ}[ \mathbf{N}^A_1 (\mathbf{x} + a \hat{x} + a \hat{y})] + \int dt \, \mathbf{M}^A_1(\mathbf{x} + a \hat{x} + a \hat{y})\cdot \left( \mathbf{N}^A_1 (\mathbf{x} + a \hat{x} + a \hat{y}) \times \partial_t \mathbf{N}^A_1 (\mathbf{x} + a \hat{x} + a \hat{y}) \right)
  \\
  &\approx S_\mathrm{WZ}[ \mathbf{N}^A_1 (\mathbf{x}) + a(\partial_x + \partial_y) \mathbf{N}^A_1(\mathbf{x})] + \int dt \, \mathbf{M}^A_1(\mathbf{x})\cdot \left( \mathbf{N}^A_1 (\mathbf{x}) \times \partial_t \mathbf{N}^A_1 (\mathbf{x}) \right)
  \\
  & \quad +\int dt \, a(\partial_x+\partial_y)\left[\mathbf{M}^A_1(\mathbf{x})\cdot \left( \mathbf{N}^A_1 (\mathbf{x}) \times \partial_t \mathbf{N}^A_1 (\mathbf{x}) \right) \right]
  \\
  & \approx S_\mathrm{WZ}[ \mathbf{N}^A_1 (\mathbf{x})] + \int dt \frac{\delta S_\mathrm{WZ}}{\delta \mathbf{S}}[\mathbf{N}_1^A(\mathbf{x})] \cdot a (\partial_x + \partial_y) \mathbf{N}_1^A(\mathbf{x}) + \int dt \, \mathbf{M}^A_1(\mathbf{x})\cdot \left( \mathbf{N}^A_1 (\mathbf{x}) \times \partial_t \mathbf{N}^A_1 (\mathbf{x}) \right)
  \\
  &=S_\mathrm{WZ}[ \mathbf{N}^A_1 (\mathbf{x})] + \int dt \left( \mathbf{N}_1^A(\mathbf{x})\times \partial_t  \mathbf{N}_1^A(\mathbf{x})\right) \cdot a (\partial_x + \partial_y) \mathbf{N}_1^A(\mathbf{x}) + \int dt \, \mathbf{M}^A_1(\mathbf{x})\cdot \left( \mathbf{N}^A_1 (\mathbf{x}) \times \partial_t \mathbf{N}^A_1 (\mathbf{x}) \right)\,.
\end{aligned}
\end{equation}
Therefore the Wess-Zumino term for layer $A$ is
\begin{equation}
\begin{aligned}
  S^A_\mathrm{WZ} &= \sum_{\mathbf{x} \in (2a \mathbb{Z})^2} \left[S_\mathrm{WZ}[\mathbf{S}_1^A(\mathbf{x})] + S_\mathrm{WZ}[\mathbf{S}_2^A(\mathbf{x} + a \hat{x})] + S_\mathrm{WZ}[\mathbf{S}_3^A(\mathbf{x} + a \hat{x} + a \hat{y})] + S_\mathrm{WZ}[\mathbf{S}_4^A(\mathbf{x} + a \hat{y})]\right]
  \\
  & = \int \frac{d^2x}{4a^2} dt \bigg[  \mathbf{N}^A_1 \cdot \left( \partial_t \mathbf{N}^A_1 \times a (\partial_x + \partial_y) \mathbf{N}^A_1 \right) + \mathbf{N}^A_2 \cdot \left( \partial_t \mathbf{N}^A_2 \times a (- \partial_x + \partial_y) \mathbf{N}^A_2 \right)
  \\
  & \quad + 2 \mathbf{M}^A_1\cdot \left( \mathbf{N}^A_1  \times \partial_t \mathbf{N}^A_1  \right) + 2 \mathbf{M}^A_2\cdot \left( \mathbf{N}^A_2  \times \partial_t \mathbf{N}^A_2  \right) \bigg]\,.
\end{aligned}
\end{equation}
The first two terms are the topological terms associated to $\pi_2(S^2)$ for $\mathbf{N}^A_1$ and $\mathbf{N}^A_2$, respectively. Hereafter the topological terms will be omitted since they do not contribute the equation of motion. Thus before integrating out the fluctuation fields, the kinetic term of layer $A$ is
\begin{align}
  S_\mathrm{WZ}^A = \int \frac{d^2x}{4a^2} dt \,2 \bigg[  \mathbf{M}^A_1\cdot \left( \mathbf{N}^A_1  \times \partial_t \mathbf{N}^A_1  \right) + \mathbf{M}^A_2\cdot \left( \mathbf{N}^A_2  \times \partial_t \mathbf{N}^A_2  \right) \bigg]\,.
\end{align}
Similarly, for layer $B$, we have
\begin{align}
  S_\mathrm{WZ}^B = \int \frac{d^2x}{4a^2} dt \,2 \bigg[  \mathbf{M}^B_1\cdot \left( \mathbf{N}^B_1  \times \partial_t \mathbf{N}^B_1  \right) + \mathbf{M}^B_2\cdot \left( \mathbf{N}^B_2  \times \partial_t \mathbf{N}^B_2  \right) \bigg]\,.
\end{align}
The leading-order action is then
\begin{equation}
\begin{aligned}
  \mathcal{S} = \int \frac{d^2x}{4a^2} dt \, &\bigg[ 2\left( \mathbf{M}^A_1\cdot \left( \mathbf{N}^A_1  \times \partial_t \mathbf{N}^A_1  \right) + \mathbf{M}^A_2\cdot \left( \mathbf{N}^A_2  \times \partial_t \mathbf{N}^A_2  \right) +  \mathbf{M}^B_1\cdot \left( \mathbf{N}^B_1  \times \partial_t \mathbf{N}^B_1  \right) + \mathbf{M}^B_2\cdot \left( \mathbf{N}^B_2  \times \partial_t \mathbf{N}^B_2  \right) \right)
  \\
  &+ 4(J_1 - J_2) (\mathbf{N}_1^A \cdot \mathbf{N}_2^A + \mathbf{N}_1^B \cdot \mathbf{N}_2^B) + 4(J_1 + J_2)(\mathbf{M}_1^A \cdot \mathbf{M}_2^A + \mathbf{M}_1^B \cdot \mathbf{M}_2^B) 
  \\
  &+ 2J_3\left( \mathbf{N}_1^A \cdot \mathbf{N}_1^B - \mathbf{N}_2^A \cdot \mathbf{N}_2^B + \mathbf{M}_1^A \cdot \mathbf{M}_1^B + \mathbf{M}_2^A \cdot \mathbf{M}_2^B \right) \bigg]\,.
\end{aligned}
\end{equation}
We expand the zeroth order term up to the second order of $\mathbf{M}^A_{1,2}$ and $\mathbf{M}^B_{1,2}$ as
\begin{equation}
\begin{aligned}
  \mathcal{S} \approx \int \frac{d^2x}{4a^2} dt \, &\bigg[ 2\left( \mathbf{M}^A_1\cdot \left( \mathbf{N}^A_1  \times \partial_t \mathbf{N}^A_1  \right) + \mathbf{M}^A_2\cdot \left( \mathbf{N}^A_2  \times \partial_t \mathbf{N}^A_2  \right) +  \mathbf{M}^B_1\cdot \left( \mathbf{N}^B_1  \times \partial_t \mathbf{N}^B_1  \right) + \mathbf{M}^B_2\cdot \left( \mathbf{N}^B_2  \times \partial_t \mathbf{N}^B_2  \right) \right)
  \\
  & -2 (J_1 - J_2)\cos(\theta_1 - \theta_2)\left(\mathbf{M}^A_1 \cdot \mathbf{M}^A_1 + \mathbf{M}^A_2 \cdot \mathbf{M}^A_2 + \mathbf{M}^B_1 \cdot \mathbf{M}^B_1 + \mathbf{M}^B_2 \cdot \mathbf{M}^B_2   \right) 
  \\
  & - J_3\cos 2\theta_1 \left( \mathbf{M}^A_1 \cdot \mathbf{M}^A_1 + \mathbf{M}^B_1 \cdot \mathbf{M}^B_1 \right) + J_3\cos 2\theta_2\left( \mathbf{M}^A_2 \cdot \mathbf{M}^A_2 + \mathbf{M}^B_2 \cdot \mathbf{M}^B_2 \right) 
  \\
  & + 4(J_1 + J_2)(\mathbf{M}_1^A \cdot \mathbf{M}_2^A + \mathbf{M}_1^B \cdot \mathbf{M}_2^B) + 2J_3\left( \mathbf{M}_1^A \cdot \mathbf{M}_1^B + \mathbf{M}_2^A \cdot \mathbf{M}_2^B \right)\bigg]
  \\
  = \int \frac{d^2x}{4a^2}dt\, &\bigg[2\left( \mathbf{M}^A_1\cdot \left( \mathbf{N}^A_1  \times \partial_t \mathbf{N}^A_1  \right) + \mathbf{M}^A_2\cdot \left( \mathbf{N}^A_2  \times \partial_t \mathbf{N}^A_2  \right) +  \mathbf{M}^B_1\cdot \left( \mathbf{N}^B_1  \times \partial_t \mathbf{N}^B_1  \right) + \mathbf{M}^B_2\cdot \left( \mathbf{N}^B_2  \times \partial_t \mathbf{N}^B_2  \right) \right)
  \\
  & -\left(2 (J_1 - J_2) \cos(\theta_1 - \theta_2) + J_3\cos 2\theta_1 \right)\left(\mathbf{M}^A_1 \cdot \mathbf{M}^A_1 + \mathbf{M}^B_1 \cdot \mathbf{M}^B_1    \right) 
  \\
  & -\left(2 (J_1 - J_2) \cos(\theta_1 - \theta_2) - J_3\cos 2\theta_2 \right)\left(\mathbf{M}^A_2 \cdot \mathbf{M}^A_2 + \mathbf{M}^B_2 \cdot \mathbf{M}^B_2    \right) 
  \\
  & + 4(J_1 + J_2)(\mathbf{M}_1^A \cdot \mathbf{M}_2^A + \mathbf{M}_1^B \cdot \mathbf{M}_2^B) + 2J_3\left( \mathbf{M}_1^A \cdot \mathbf{M}_1^B + \mathbf{M}_2^A \cdot \mathbf{M}_2^B \right) \bigg]
  \\
  = \int \frac{d^2x}{4a^2}dt\, &\bigg[2\left( \mathbf{M}^A_1\cdot \left( \mathbf{N}^A_1  \times \partial_t \mathbf{N}^A_1  \right) + \mathbf{M}^A_2\cdot \left( \mathbf{N}^A_2  \times \partial_t \mathbf{N}^A_2  \right) +  \mathbf{M}^B_1\cdot \left( \mathbf{N}^B_1  \times \partial_t \mathbf{N}^B_1  \right) + \mathbf{M}^B_2\cdot \left( \mathbf{N}^B_2  \times \partial_t \mathbf{N}^B_2  \right) \right)
  \\
  & -\left(2 (J_1 - J_2) \sin 2\theta_2 - J_3\cos 2\theta_2 \right)\left(\mathbf{M}^A_1 \cdot \mathbf{M}^A_1 + \mathbf{M}^B_1 \cdot \mathbf{M}^B_1  + \mathbf{M}^A_2 \cdot \mathbf{M}^A_2 + \mathbf{M}^B_2 \cdot \mathbf{M}^B_2     \right) 
  \\
  & + 4(J_1 + J_2)(\mathbf{M}_1^A \cdot \mathbf{M}_2^A + \mathbf{M}_1^B \cdot \mathbf{M}_2^B) + 2J_3\left( \mathbf{M}_1^A \cdot \mathbf{M}_1^B + \mathbf{M}_2^A \cdot \mathbf{M}_2^B \right) \bigg]
  \\
  = \int \frac{d^2x}{4a^2}dt\, &\bigg[2\left( \mathbf{M}^A_1\cdot \left( \mathbf{N}^A_1  \times \partial_t \mathbf{N}^A_1  \right) + \mathbf{M}^A_2\cdot \left( \mathbf{N}^A_2  \times \partial_t \mathbf{N}^A_2  \right) +  \mathbf{M}^B_1\cdot \left( \mathbf{N}^B_1  \times \partial_t \mathbf{N}^B_1  \right) + \mathbf{M}^B_2\cdot \left( \mathbf{N}^B_2  \times \partial_t \mathbf{N}^B_2  \right) \right)
  \\
  & -\sqrt{4(J_1 - J_2)^2 + J_3^2}\left(\mathbf{M}^A_1 \cdot \mathbf{M}^A_1 + \mathbf{M}^B_1 \cdot \mathbf{M}^B_1  + \mathbf{M}^A_2 \cdot \mathbf{M}^A_2 + \mathbf{M}^B_2 \cdot \mathbf{M}^B_2     \right) 
  \\
  & + 4(J_1 + J_2)(\mathbf{M}_1^A \cdot \mathbf{M}_2^A + \mathbf{M}_1^B \cdot \mathbf{M}_2^B) + 2J_3\left( \mathbf{M}_1^A \cdot \mathbf{M}_1^B + \mathbf{M}_2^A \cdot \mathbf{M}_2^B \right) \bigg]\,.
\end{aligned}
\end{equation}
To obtain the effective action written interms of the N\'eel fields, we integrate out the fluctuation $\mathcal{M} \equiv (\mathbf{M}^A_1, \mathrm{sgn}(J_1 + J_2)\mathbf{M}^A_2, \mathrm{sgn}(J_3)\mathbf{M}^B_1, \mathrm{sgn}(J_1 + J_2)\mathrm{sgn}(J_3)\mathbf{M}^B_2)$ by considering the following path integral
\begin{align}
  \mathcal{Z} &= \int \mathcal{D} \mathcal{N} \mathcal{D} \mathcal{M} \delta(|\mathbf{N}^{A,B}_{1,2}|^2 - 1)\delta(\mathbf{M}^{A,B}_{1,2} \cdot \mathbf{N}^{A,B}_{1,2}) \exp\left[i \mathcal{S}\right] \nonumber
  \\
  &\propto \int \mathcal{D} \mathcal{N} \mathcal{D} \mathcal{M} \mathcal{D} \lambda \delta(|\mathbf{N}^{A,B}_{1,2}|^2 - 1) \exp\left[i \int \frac{d^2 x}{4a^2} dt \left( -\frac{1}{2} \mathcal{M} \cdot \mathcal{J} \mathcal{M} + (2 \dot{\mathcal{N}} + \Lambda \mathcal{N})\cdot \mathcal{M} \right)\right]\,,
\end{align}
where $\mathcal{N} \equiv (\mathbf{N}^A_1, \mathrm{sgn}(J_1 + J_2)\mathbf{N}^A_2, \mathrm{sgn}(J_3)\mathbf{N}^B_1, \mathrm{sgn}(J_1 + J_2)\mathrm{sgn}(J_3) \mathbf{N}^B_2)$, $\dot{\mathcal{N}} \equiv ( \mathbf{N}^A_1 \times \partial_t \mathbf{N}^A_1,\mathrm{sgn}(J_1 + J_2) \mathbf{N}^A_2 \times \partial_t \mathbf{N}^A_2, \mathrm{sgn}(J_3)\mathbf{N}^B_1 \times \partial_t \mathbf{N}^B_1, \mathrm{sgn}(J_1 + J_2)\mathrm{sgn}(J_3)\mathbf{N}^B_2 \times \partial_t \mathbf{N}^B_2)$, $\Lambda = \mathrm{diag}(\lambda^A_1, \lambda^A_2, \lambda^B_1, \lambda^B_2)$, $\lambda^T = (\lambda^A_1, \lambda^A_2, \lambda^B_1, \lambda^B_2)$, and
\begin{align}
  \mathcal{J} &= \begin{pmatrix}
    2\sqrt{4(J_1 - J_2)^2 + J_3^2} & -4|J_1 + J_2| & -2|J_3| & 0 \\
    -4|J_1 + J_2| & 2\sqrt{4(J_1 - J_2)^2 + J_3^2} & 0 & -2|J_3| \\
    -2|J_3| & 0 & 2\sqrt{4(J_1 - J_2)^2 + J_3^2} & -4|J_1 + J_2| \\
    0 & -2|J_3| & -4|J_1 + J_2| & 2\sqrt{4(J_1 - J_2)^2 + J_3^2}
  \end{pmatrix}\,.
\end{align}
From the change of variavles according to $\mathcal{M} = \tilde{\mathcal{M}} + \mathcal{J}^{-1} (2\dot{\mathcal{N}} + \Lambda \mathcal{N})$, we have
\begin{align}
  \mathcal{Z} &\propto \int \mathcal{D} \mathcal{N} \mathcal{D} \tilde{\mathcal{M}} \mathcal{D} \lambda \delta(|\mathbf{N}^{A,B}_{1,2}|^2 - 1) \exp\left[i \int \frac{d^2 x}{4a^2} dt \left( -\frac{1}{2} \tilde{\mathcal{M}} \cdot \mathcal{J} \tilde{\mathcal{M}} + \frac{1}{2} (2\dot{\mathcal{N}} + \Lambda \mathcal{N})\cdot \mathcal{J}^{-1} (2\dot{\mathcal{N}} + \Lambda \mathcal{N}) \right)\right] \nonumber
  \\
  &\propto \int \mathcal{D} \mathcal{N}  \mathcal{D} \lambda \delta(|\mathbf{N}^{A,B}_{1,2}|^2 - 1) \exp\left[i \int \frac{d^2 x}{4a^2} dt \left( 2 \dot{\mathcal{N}} \cdot \mathcal{J}^{-1} \dot{\mathcal{N}} + 2 \dot{\mathcal{N}} \cdot\mathcal{J}^{-1} \Lambda \mathcal{N} + \frac{1}{2} \Lambda \mathcal{N} \cdot \mathcal{J}^{-1} \Lambda \mathcal{N} \right)\right]
  \\
  &= \int \mathcal{D} \mathcal{N}  \mathcal{D} \lambda \delta(|\mathbf{N}^{A,B}_{1,2}|^2 - 1) \exp\left[i \int \frac{d^2 x}{4a^2} dt \left( 2 \dot{\mathcal{N}} \cdot \mathcal{J}^{-1} \dot{\mathcal{N}} +j^T \lambda + \frac{1}{2} \lambda^T A \lambda \right)\right]\,, \nonumber
\end{align}
where
\begin{align}
  \mathcal{J}^{-1} &= \frac{1}{\mathcal{A}}\begin{pmatrix}
    \frac{1}{8}J_1 J_2 \mathcal{E}_{nc} & \frac{1}{8}|J_1 + J_2|(J_3^2 -8 J_1 J_2) & \frac{1}{4}(J_1^2 + J_2^2)|J_3| & -\frac{1}{32} \mathcal{E}_{nc}|J_1 + J_2||J_3| \\
    \frac{1}{8}|J_1 + J_2|(J_3^2 -8 J_1 J_2) & \frac{1}{8}J_1 J_2 \mathcal{E}_{nc} & -\frac{1}{32} \mathcal{E}_{nc}|J_1 + J_2||J_3| & \frac{1}{4}(J_1^2 + J_2^2)|J_3| \\
    \frac{1}{4}(J_1^2 + J_2^2)|J_3| & -\frac{1}{32} \mathcal{E}_{nc}|J_1 + J_2||J_3| & \frac{1}{8}J_1 J_2 \mathcal{E}_{nc} & \frac{1}{8}|J_1 + J_2|(J_3^2 -8 J_1 J_2) \\
    -\frac{1}{32} \mathcal{E}_{nc}|J_1 + J_2||J_3| & \frac{1}{4}(J_1^2 + J_2^2)|J_3| & \frac{1}{8}|J_1 + J_2|(J_3^2 -8 J_1 J_2) & \frac{1}{8}J_1 J_2 \mathcal{E}_{nc}
  \end{pmatrix}\,,  \nonumber
  \\
  \mathcal{A} &= (4J_1 J_2)^2 - (J_1 + J_2)^2 J_3^2\,,
\end{align}
and
\begin{align}
  A &= \begin{pmatrix}
    [\mathcal{J}^{-1}]_{11} & [A]_{12} & [A]_{13} & 0 \\
    [A]_{12} & [\mathcal{J}^{-1}]_{11} & 0 & -[A]_{13} \\
    [A]_{13} & 0 & [\mathcal{J}^{-1}]_{11} & [A]_{12} \\
    0 & -[A]_{13} & [A]_{12} & [\mathcal{J}^{-1}]_{11}
  \end{pmatrix}\,, \nonumber
  \\
  [A]_{12} &= [\mathcal{J}^{-1}]_{12} \mathrm{sgn}(J_1 + J_2) \cos(\theta_1 - \theta_2) = [\mathcal{J}^{-1}]_{12} \mathrm{sgn}(J_1 + J_2) \sin 2\theta_2\,, \nonumber
  \\
  [A]_{13} &= [\mathcal{J}^{-1}]_{13}\mathrm{sgn}(J_3)\cos 2\theta_1 = - [\mathcal{J}^{-1}]_{13}\mathrm{sgn}(J_3)\cos 2\theta_2 \,, \nonumber
  \\
  j & = 2\begin{pmatrix}
    -\frac{1}{4}\rho_\mathrm{intra}\sin(\theta_1 - \theta_2) \hat{\gamma} \cdot \partial_t \mathbf{N}^A_2 -\frac{1}{4}\rho_\mathrm{inter}\sin(2\theta_1) \hat{\gamma} \cdot \partial_t \mathbf{N}^B_1 -\frac{1}{4}\rho_\mathrm{ortho}\hat{\gamma} \cdot \partial_t \mathbf{N}^B_2 \\
    \frac{1}{4}\rho_\mathrm{intra}\sin(\theta_1 - \theta_2) \hat{\gamma} \cdot \partial_t \mathbf{N}^A_1 -\frac{1}{4}\rho_\mathrm{inter}\sin(2\theta_2) \hat{\gamma} \cdot \partial_t \mathbf{N}^B_2 -\frac{1}{4}\rho_\mathrm{ortho} \hat{\gamma} \cdot \partial_t \mathbf{N}^B_1 \\
    \frac{1}{4}\rho_\mathrm{intra}\sin(\theta_1 - \theta_2) \hat{\gamma} \cdot \partial_t \mathbf{N}^B_2 + \frac{1}{4}\rho_\mathrm{inter}\sin(2\theta_1) \hat{\gamma} \cdot \partial_t \mathbf{N}^A_1 + \frac{1}{4}\rho_\mathrm{ortho} \hat{\gamma} \cdot \partial_t \mathbf{N}^A_2  \\
    -\frac{1}{4}\rho_\mathrm{intra}\sin(\theta_1 - \theta_2) \hat{\gamma} \cdot \partial_t \mathbf{N}^B_1 + \frac{1}{4}\rho_\mathrm{inter}\sin(2\theta_2) \hat{\gamma} \cdot \partial_t \mathbf{N}^A_2 + \frac{1}{4}\rho_\mathrm{ortho} \hat{\gamma} \cdot \partial_t \mathbf{N}^A_1
  \end{pmatrix}\,,
  \\
  \rho_\mathrm{diag} &= 2\left[ \mathcal{J}^{-1}\right]_{11} = \frac{\mathcal{E}_{nc}J_1 J_2}{4\left[(4J_1 J_2)^2 - (J_1 + J_2)^2 J_3^2\right]}\,, \nonumber
  \\
  \rho_\mathrm{intra} &= 4\left[ \mathcal{J}^{-1}\right]_{12} \mathrm{sgn}(J_1 + J_2)= \frac{(J_1 + J_2)\left(J_3^2 -8 J_1 J_2   \right)}{2\left[(4J_1 J_2)^2 - (J_1 + J_2)^2 J_3^2\right]}\,, \nonumber
  \\
  \rho_\mathrm{inter} &= 4\left[ \mathcal{J}^{-1}\right]_{13} \mathrm{sgn}(J_3) = \frac{\left(J_1^2 + J_2^2\right)J_3}{\left[(4J_1 J_2)^2 - (J_1 + J_2)^2 J_3^2\right]}  \,, \nonumber
  \\
  \rho_\mathrm{ortho} &= 4\left[ \mathcal{J}^{-1}\right]_{14} \mathrm{sgn}(J_1 + J_2)\mathrm{sgn}(J_3)= -\frac{\mathcal{E}_{nc}\left(J_1 + J_2\right)J_3}{8\left[(4J_1 J_2)^2 - (J_1 + J_2)^2 J_3^2\right]}  \,, \nonumber
\end{align}
with the axis $ \hat{\gamma} = \mathbf{N}^B_2 \times \mathbf{N}^A_1 $ perpendicular to the plane defined by the coplanar spin order . Performing the Gaussian integral with $\lambda = \tilde{\lambda} - A^{-1} j$, we obtain the effective action for the N\'eel fields
\begin{align}
  \mathcal{Z} &\propto \int \mathcal{D} \mathcal{N}  \mathcal{D} \tilde{\lambda} \delta(|\mathbf{N}^{A,B}_{1,2}|^2 - 1) \exp\left[i \int \frac{d^2 x}{4a^2} dt \left( 2 \dot{\mathcal{N}} \cdot \mathcal{J}^{-1} \dot{\mathcal{N}} - \frac{1}{2} j^T A^{-1} j  + \frac{1}{2} \tilde{\lambda}^T A \tilde{\lambda}\right)\right] \nonumber
  \\
  &\propto \int \mathcal{D} \mathcal{N}  \delta(|\mathbf{N}^{A,B}_{1,2}|^2 - 1) \exp\left[i \int \frac{d^2 x}{4a^2} dt \left( 2 \dot{\mathcal{N}} \cdot \mathcal{J}^{-1} \dot{\mathcal{N}} - \frac{1}{2} j^T A^{-1} j \right)\right]\,.
\end{align}
Note that, in the weak-coupling limit, $j$ is negligible and the path integral becomes
\begin{align}
  \mathcal{Z} \sim \int \mathcal{D} \mathcal{N}  \delta(|\mathbf{N}^{A,B}_{1,2}|^2 - 1) \exp\left[i \int \frac{d^2 x}{4a^2} dt \left( 2 \dot{\mathcal{N}} \cdot \mathcal{J}^{-1} \dot{\mathcal{N}} \right)\right]\,.
\end{align}
After integrating out the magnetization fields and the auxiliary fields for the constraints, we obtain the effective kinetic term for the N\'eel fields
\begin{equation}
\begin{aligned}
  S_\mathrm{kin} = \int \frac{d^2x}{4a^2} dt \, &\bigg[ \rho_\mathrm{diag}\left(\left|\mathbf{N}^A_1 \times \partial_t \mathbf{N}^A_1 \right|^2 + \left|\mathbf{N}^A_2 \times \partial_t \mathbf{N}^A_2 \right|^2 + \left|\mathbf{N}^B_1 \times \partial_t \mathbf{N}^B_1 \right|^2 + \left|\mathbf{N}^B_2 \times \partial_t \mathbf{N}^B_2 \right|^2 \right)
  \\
  & +  \rho_\mathrm{intra}\left( \left(\mathbf{N}^A_1 \times \partial_t \mathbf{N}^A_1 \right)\cdot\left(\mathbf{N}^A_2 \times \partial_t \mathbf{N}^A_2 \right) + \left(\mathbf{N}^B_1 \times \partial_t \mathbf{N}^B_1 \right)\cdot\left(\mathbf{N}^B_2 \times \partial_t \mathbf{N}^B_2 \right)\right)
  \\
  & + \rho_\mathrm{inter}\left( \left(\mathbf{N}^A_1 \times \partial_t \mathbf{N}^A_1 \right)\cdot\left(\mathbf{N}^B_1 \times \partial_t \mathbf{N}^B_1 \right) + \left(\mathbf{N}^A_2 \times \partial_t \mathbf{N}^A_2 \right)\cdot\left(\mathbf{N}^B_2 \times \partial_t \mathbf{N}^B_2 \right)\right)
  \\
  & + \rho_\mathrm{ortho}\left(\left(\mathbf{N}^A_1\times \mathbf{N}^A_1\right)\cdot\left(\mathbf{N}^B_2\times \mathbf{N}^B_2\right) + \left(\mathbf{N}^A_2\times \mathbf{N}^A_2\right)\cdot\left(\mathbf{N}^B_1\times \mathbf{N}^B_1\right)\right) - \frac{1}{2} j^T A^{-1} j\bigg]\,,
\end{aligned}
\end{equation}
where
\begin{equation}
\begin{aligned}
  \rho_\mathrm{diag} &=  \frac{\mathcal{E}_{nc}J_1 J_2}{4\left[(4J_1 J_2)^2 - (J_1 + J_2)^2 J_3^2\right]}\,,
  \\
  \rho_\mathrm{intra} &=  \frac{(J_1 + J_2)\left(J_3^2 -8 J_1 J_2   \right)}{2\left[(4J_1 J_2)^2 - (J_1 + J_2)^2 J_3^2\right]}\,,
  \\
  \rho_\mathrm{inter} &=  \frac{\left(J_1^2 + J_2^2\right)J_3}{\left[(4J_1 J_2)^2 - (J_1 + J_2)^2 J_3^2\right]}  \,,
  \\
  \rho_\mathrm{ortho} &=  -\frac{\mathcal{E}_{nc}\left(J_1 + J_2\right)J_3}{8\left[(4J_1 J_2)^2 - (J_1 + J_2)^2 J_3^2\right]}  \,.
\end{aligned}
\end{equation}
The kinetic term can be reexpressed as
\begin{equation}
\begin{aligned}
  S_\mathrm{kin} = \int \frac{d^2x}{4a^2} dt \, & \bigg[ \rho_\mathrm{diag} \left( \left|\partial_t\mathbf{N}^A_1\right|^2 + \left|\partial_t\mathbf{N}^A_2\right|^2 + \left|\partial_t\mathbf{N}^B_1\right|^2 + \left|\partial_t\mathbf{N}^B_2\right|^2 \right)
  \\
  & + \rho_\mathrm{intra} \left(\cos(\theta_1 - \theta_2) \partial_t \mathbf{N}^A_1 \cdot \partial_t \mathbf{N}^A_2 + \left( \mathbf{N}^A_1 \cdot \partial_t \mathbf{N}^A_2 \right)^2 \right)
  \\
  & + \rho_\mathrm{intra} \left(\cos(\theta_1 - \theta_2) \partial_t \mathbf{N}^B_1 \cdot \partial_t \mathbf{N}^B_2 + \left( \mathbf{N}^B_1 \cdot \partial_t \mathbf{N}^B_2 \right)^2 \right)
  \\
  & + \rho_\mathrm{inter} \left(\cos2\theta_1 \partial_t \mathbf{N}^A_1 \cdot \partial_t \mathbf{N}^B_1  + \left( \mathbf{N}^A_1 \cdot \partial_t \mathbf{N}^B_1 \right)^2 \right)
  \\
  & + \rho_\mathrm{inter} \left(\cos2\theta_2 \partial_t \mathbf{N}^A_2 \cdot \partial_t \mathbf{N}^B_2  + \left( \mathbf{N}^A_2 \cdot \partial_t \mathbf{N}^B_2 \right)^2 \right)
  \\
  & + \rho_\mathrm{ortho} \left(\left( \mathbf{N}^A_1 \cdot \partial_t \mathbf{N}^B_2 \right)^2 + \left( \mathbf{N}^A_2 \cdot \partial_t \mathbf{N}^B_1 \right)^2 \right) - \frac{1}{2} j^T A^{-1} j \bigg]\,.
\end{aligned}
\end{equation}
Finally, in terms of the $\mathrm{SO}(3)$ field, the kinetic term becomes
\begin{equation}
\begin{aligned}
  S_\mathrm{kin} = \int \frac{d^2x}{4a^2} dt \, &\bigg[ \rho_\mathrm{diag} \mathrm{Tr}\left[ \partial_t \mathcal{R}^T \partial_t \mathcal{R}  \left(|n^A_1 \rangle \langle n^A_1| + |n^A_2\rangle \langle n^A_2| + |n^B_1 \rangle \langle n^B_1| +|n^B_2 \rangle \langle n^B_2|\right) \right] 
  \\
  & + \rho_\mathrm{intra} \left(\sin 2\theta_2 \mathrm{Tr}\left[ \partial_t \mathcal{R}^T \partial_t \mathcal{R} |n^A_2 \rangle \langle n^A_1| \right] + \mathrm{Tr}\left[ \mathcal{R}^T \partial_t \mathcal{R} | n^A_2 \rangle \langle n^A_1 | \right]^2 \right)
  \\
  & + \rho_\mathrm{intra} \left(\sin 2\theta_2 \mathrm{Tr}\left[ \partial_t \mathcal{R}^T \partial_t \mathcal{R} |n^B_2 \rangle \langle n^B_1| \right] + \mathrm{Tr}\left[ \mathcal{R}^T \partial_t \mathcal{R} | n^B_2 \rangle \langle n^B_1 | \right]^2 \right)
  \\
  & + \rho_\mathrm{inter} \left(-\cos2\theta_2 \mathrm{Tr}\left[ \partial_t \mathcal{R}^T \partial_t \mathcal{R} |n^B_1 \rangle \langle n^A_1| \right] + \mathrm{Tr}\left[ \mathcal{R}^T \partial_t \mathcal{R} | n^B_1 \rangle \langle n^A_1 | \right]^2 \right)
  \\
  & + \rho_\mathrm{inter} \left(\cos2\theta_2 \mathrm{Tr}\left[ \partial_t \mathcal{R}^T \partial_t \mathcal{R} |n^B_2 \rangle \langle n^A_2| \right] + \mathrm{Tr}\left[ \mathcal{R}^T \partial_t \mathcal{R} | n^B_2 \rangle \langle n^A_2 | \right]^2 \right)
  \\
  & + \rho_\mathrm{ortho} \left(\mathrm{Tr}\left[ \mathcal{R}^T \partial_t \mathcal{R} | n^B_2 \rangle \langle n^A_1 | \right]^2 + \mathrm{Tr}\left[ \mathcal{R}^T \partial_t \mathcal{R} | n^B_1 \rangle \langle n^A_2 | \right]^2 \right) - \frac{1}{2} j^T A^{-1} j \bigg]\,.
\end{aligned}
\end{equation}
We can recast the above action into the following compact form:
\begin{equation}
\begin{aligned}
    S_\mathrm{kin} = \int \frac{d^2x}{4a^2} dt \sum_{IJij}\bigg[&C^{IJ}_{ij} \mathrm{Tr} \left[\partial_t \mathcal{R}^T \partial_t \mathcal{R} |n^I_i\rangle \langle n^J_j| \right]
    + D^{IJ}_{ij} \mathrm{Tr} \left[ \mathcal{R}^T \partial_t \mathcal{R} |n^I_i\rangle \langle n^J_j| \right]^2 
    \\
    &+ F^{IJ}_{ij} \mathrm{Tr}\left[ \mathcal{R}^T\partial_t \mathcal{R} |n^I_i\rangle \langle \gamma_0|\right]\mathrm{Tr}\left[\mathcal{R}^T\partial_t \mathcal{R} |n^J_j\rangle \langle \gamma_0|\right] \bigg]\,,
\end{aligned}
\end{equation}
where $\gamma_0 = n^B_2\times n^A_1$.
Here, the coefficients $C^{IJ}_{ij}$, $D^{IJ}_{ij}$, and $F^{IJ}_{ij}$, which are symmetric under the exchange of indices $(I \leftrightarrow J)$ and $(i \leftrightarrow j)$, are given by $C^{II}_{ii} = \rho_\mathrm{diag}$, $C^{AA}_{12} = C^{BB}_{12}=\rho_\mathrm{intra} \sin 2\theta_2/2$, $C^{AB}_{22} = -C^{AB}_{11} = \rho_\mathrm{inter}\cos 2\theta_2/2$, $C^{AB}_{12}=0$, $D^{II}_{ii}=0$, $D^{AA}_{12} = D^{BB}_{12} = \rho_\mathrm{intra}/2$, $D^{AB}_{22} = D^{AB}_{11} = \rho_\mathrm{inter}/2$, $D^{AB}_{12} = \rho_\mathrm{ortho}$,
$F^{II}_{ii} = \mathcal{F}\mathcal{E}_{nc}^2 ( J_3^2 (4 (J_1^4 - 2 J_1^3 J_2 - 2 J_1^2 J_2^2 - 2 J_1 J_2^3 + J_2^4) + (J_1 + J_2)^2 J_3^2))/128  $, $F^{AB}_{12} = 0$,
$F^{AA}_{12} = F^{BB}_{12} = \mathcal{F} ((J_1^4 - J_2^4) J_3^2 (2 (J_1^2 - 4 J_1 J_2 + J_2^2) + J_3^2))$, and
$F^{AB}_{11} = - F^{AB}_{22} = \mathcal{F} (-4 J_1 J_2 (J_1^2 - J_2^2)^2 J_3^2 + (J_1^4 - J_1^3 J_2 - 2 J_1^2 J_2^2 - J_1 J_2^3 + J_2^4) J_3^4 +  (J_1 + J_2)^2 J_3^6/8)$, where $\mathcal{F} = 1/[\mathcal{E}_{nc} (J_1-J_2)^2 (J_3^2 - 8 J_1 J_2)[(4J_1 J_2)^2 -(J_1+J_2)^2 J_3^2]]$.

\subsection{Collinear antiferromagnetic phase}
Now we consider the collinear antiferromagnetic phases. From Eq.~\eqref{eq:Swz}, we have
\begin{equation}
\begin{aligned}
  S_\mathrm{WZ}[\mathbf{S}^A_1(\mathbf{x})] &= S_\mathrm{WZ}[ -\mathbf{N}^A_1 (\mathbf{x}) + \mathbf{M}^A_1(\mathbf{x})]
  \\
  &\approx S_\mathrm{WZ}[\mathbf{M}^A_1 (\mathbf{x})] + \int dt\, \mathbf{N}^A_1(\mathbf{x})\cdot \left( \mathbf{M}^A_1 (\mathbf{x}) \times \partial_t \mathbf{M}^A_1 (\mathbf{x}) \right)
\end{aligned}
\end{equation}
and therefore
\begin{equation}
\begin{aligned}
  S^A_\mathrm{WZ} &= \sum_{\mathbf{x} \in (2a \mathbb{Z})^2} \left[S_\mathrm{WZ}[\mathbf{S}_1^A(\mathbf{x})] + S_\mathrm{WZ}[\mathbf{S}_2^A(\mathbf{x} + a \hat{x})] + S_\mathrm{WZ}[\mathbf{S}_3^A(\mathbf{x} + a \hat{x} + a \hat{y})] + S_\mathrm{WZ}[\mathbf{S}_4^A(\mathbf{x} + a \hat{y})]\right]
  \\
  &\approx\sum_{\mathbf{x} \in (2a \mathbb{Z})^2} 2\left[S_\mathrm{WZ}[\mathbf{M}^A_1 (\mathbf{x})] +  S_\mathrm{WZ}[\mathbf{M}^A_2 (\mathbf{x})] \right]\,.
\end{aligned}
\end{equation}
Similarly, for layer $B$, we have
\begin{equation}
\begin{aligned}
  S^B_\mathrm{WZ} &= \sum_{\mathbf{x} \in (2a \mathbb{Z})^2} \left[S_\mathrm{WZ}[\mathbf{S}_1^B(\mathbf{x})] + S_\mathrm{WZ}[\mathbf{S}_2^B(\mathbf{x} + a \hat{x})] + S_\mathrm{WZ}[\mathbf{S}_3^B(\mathbf{x} + a \hat{x} + a \hat{y})] + S_\mathrm{WZ}[\mathbf{S}_4^B(\mathbf{x} + a \hat{y})]\right]
  \\
  &\approx\sum_{\mathbf{x} \in (2a \mathbb{Z})^2} 2\left[S_\mathrm{WZ}[\mathbf{M}^B_1 (\mathbf{x})] +  S_\mathrm{WZ}[\mathbf{M}^B_2 (\mathbf{x})] \right]\,.
\end{aligned}
\end{equation}
In the AFM~I phase, using the parameterization~\eqref{eq:afm1}, the total Wess-Zumino term is
\begin{equation}
\begin{aligned}
  S^A_\mathrm{WZ} + S^B_\mathrm{WZ} &\approx \int \frac{d^2x}{2a^2} \left[ S_\mathrm{WZ} [\mathbf{M}^A_1] + S_\mathrm{WZ} [\mathbf{M}^A_2] + S_\mathrm{WZ} [\mathbf{M}^B_1] + S_\mathrm{WZ} [\mathbf{M}^B_2] \right]
  \\
  &\approx \int \frac{d^2x}{a^2} \left[S_\mathrm{WZ} [\mathbf{n} + \mathbf{m}] + S_\mathrm{WZ} [-\mathbf{n} + \mathbf{m}] \right]
  \\
  &\approx
  \int \frac{d^2x}{a^2} \left[S_\mathrm{WZ} [\mathbf{n}] + \int dt\, \mathbf{m}\cdot \left( \mathbf{n}  \times \partial_t \mathbf{n} \right)- S_\mathrm{WZ} [\mathbf{n}] + \int dt\, \mathbf{m}\cdot \left( \mathbf{n}  \times \partial_t \mathbf{n} \right)\right]
  \\
  &=
  \int \frac{d^2x}{a^2} dt\, 2\mathbf{m}\cdot \left( \mathbf{n}  \times \partial_t \mathbf{n} \right)\,.
\end{aligned}
\end{equation}
In the AFM~II phase, with the parameterization~\eqref{eq:afm2}, the sum of the Wess-Zumino terms yields the same result:
\begin{equation}
\begin{aligned}
  S^A_\mathrm{WZ} + S^B_\mathrm{WZ} &\approx \int \frac{d^2x}{2a^2} \left[ S_\mathrm{WZ} [\mathbf{M}^A_1] + S_\mathrm{WZ} [\mathbf{M}^A_2] + S_\mathrm{WZ} [\mathbf{M}^B_1] + S_\mathrm{WZ} [\mathbf{M}^B_2] \right]
  \\
  &\approx \int \frac{d^2x}{a^2} \left[S_\mathrm{WZ} [\mathbf{n} + \mathbf{m}] + S_\mathrm{WZ} [-\mathbf{n} + \mathbf{m}] \right]
  \\
  &\approx
  \int \frac{d^2x}{a^2} \left[S_\mathrm{WZ} [\mathbf{n}] + \int dt\, \mathbf{m}\cdot \left( \mathbf{n}  \times \partial_t \mathbf{n} \right)- S_\mathrm{WZ} [\mathbf{n}] + \int dt\, \mathbf{m}\cdot \left( \mathbf{n}  \times \partial_t \mathbf{n} \right)\right]
  \\
  &=
  \int \frac{d^2x}{a^2} dt\, 2\mathbf{m}\cdot \left( \mathbf{n}  \times \partial_t \mathbf{n} \right)\,.
\end{aligned}
\end{equation}
The same result holds for the AFM~III phase.

\subsection{Collinear ferromagnetic phase}

In the collinear ferromagnetic phase, the kinetic term is
\begin{align}
  S^A_\mathrm{WZ} + S^B_\mathrm{WZ} &\approx \int \frac{d^2x}{a^2} \left[ S_\mathrm{WZ} [\mathbf{M}^A] +S_\mathrm{WZ} [\mathbf{M}^B]  \right]\,.
\end{align}

\section{Actions for the collinear phases}\label{supp:col}

In this section, we summarize the actions for the collinear phases.

The action for the AFM~I phase is
\begin{align}
  S_\mathrm{AFMI} = \int d^2 x dt \left[ \frac{2}{a^2} \mathbf{m} \cdot \left( \mathbf{n} \times \partial_t \mathbf{n} \right) + \frac{2J_3}{a^2}|\mathbf{m}|^2 + (J_1 - J_2) \mathbf{m}\cdot\left(\partial_x^2 - \partial_y^2 \right) \mathbf{n} - \frac{J_1 + J_2}{2} \sum_i |\partial_i \mathbf{n}|^2 \right]\,,
\end{align}
where
\begin{align}
  \mathbf{M}^A_1 = \mathbf{M}^A_2 = \mathbf{n} + \mathbf{m}\,, \quad \mathbf{M}^B_1 = \mathbf{M}^B_2 = -\mathbf{n} + \mathbf{m}\,.
\end{align}
The action for the AFM~II phase is
\begin{align}
  S_\mathrm{AFMII} = \int d^2 x dt \left[ \frac{2}{a^2} \mathbf{m} \cdot \left( \mathbf{n} \times \partial_t \mathbf{n} \right) + \frac{4(J_1 + J_2) + 2J_3}{a^2}|\mathbf{m}|^2 - (J_1 + J_2) \mathbf{m}\cdot \sum_i \partial_i^2 \mathbf{n} + \frac{J_1 + J_2}{2} \sum_i |\partial_i \mathbf{n}|^2 \right]\,,
\end{align}
where
\begin{align}
  \mathbf{M}^A_1 = \mathbf{M}^B_2 = \mathbf{n} + \mathbf{m}\,, \quad \mathbf{M}^A_2 = \mathbf{M}^B_1 = -\mathbf{n} + \mathbf{m}\,.
\end{align}
The action for the AFM~III phase is
\begin{align}
  S_\mathrm{AFMIII} = \int d^2 x dt \left[ \frac{2}{a^2} \mathbf{m} \cdot \left( \mathbf{n} \times \partial_t \mathbf{n} \right) + \frac{4(J_1 + J_2)}{a^2}|\mathbf{m}|^2 - (J_1 + J_2) \mathbf{m}\cdot \sum_i \partial_i^2 \mathbf{n} + \frac{J_1 + J_2}{2} \sum_i |\partial_i \mathbf{n}|^2 \right]\,,
\end{align}
where
\begin{align}
  \mathbf{M}^A_1 = \mathbf{M}^B_1 = \mathbf{n} + \mathbf{m}\,, \quad \mathbf{M}^A_2 = \mathbf{M}^B_2 = -\mathbf{n} + \mathbf{m}\,.
\end{align}
The action for the FM phase is
\begin{equation}
\begin{aligned}
  S_\mathrm{FM} &= \int \frac{d^2x}{a^2} \left[ S_\mathrm{WZ}\left[\mathbf{M}^A\right] + S_\mathrm{WZ}\left[\mathbf{M}^B\right]\right] 
  \\
  & \quad + \int d^2x dt \left[ \frac{J_3}{a^2} \mathbf{M}^A \cdot \mathbf{M}^B -\left(\frac{J_1}{2} \left|\partial_x\mathbf{M}^A\right|^2 + \frac{J_2}{2} \left|\partial_y\mathbf{M}^A\right|^2 + \frac{J_2}{2} \left|\partial_x\mathbf{M}^B\right|^2 + \frac{J_1}{2} \left|\partial_y\mathbf{M}^B\right|^2 \right)\right]
\end{aligned}
\end{equation}
where
\begin{align}
  \mathbf{M} = \mathbf{M}^A_1 = \mathbf{M}^A_2\,, \quad \mathbf{M}^B = \mathbf{M}^B_1 = \mathbf{M}^B_2\,.
\end{align}

\section{Numerical Method}\label{supp:numerical}

In the region $-1 \le J_2/J_1 < 0$, we first perform simulated annealing.  Starting from a randomly generated initial configuration at inverse temperature $\beta = 0.1$, we increase $\beta$ by a factor of 1.01 every $10^4$ Monte Carlo steps, for a total of $10^7$ steps.  After annealing, we apply the gradient‐descent method known as arrested Newton flow, using the annealed configuration as the initial condition.  This combined protocol is faster than simulated annealing alone and remains unbiased with respect to the choice of initial state. In the region $J_2/J_1 < -1$, we perform similar protocols at one point in the AFM II phase and one point in the noncollinear phase. Using these two points as initial states, we perform the arrested Newton flow and obtain two minima. We compare the energies of the two minima and determine the ground state. The noncollinearity in Fig.~2 of the main text is the average of the square of the N\'eel fields.

\end{document}